\documentclass[12pt]{article}

\usepackage{amsfonts}
\usepackage{amssymb,amsmath}
\usepackage[totalwidth=460truept,totalheight=625truept]{geometry}
\usepackage{latexsym,graphicx}
\usepackage[hidelinks]{hyperref}

\global\arraycolsep=1truept

\begin{document}

\null

\vskip.6truecm

\begin{center}
{\huge \textbf{Quantum Field Theories of}}

\vskip.8truecm

{\huge \textbf{Arbitrary-Spin Massive\ Multiplets }}

\vskip.8truecm

{\huge \textbf{and Palatini Quantum Gravity}}

\vskip1truecm

\textsl{Damiano Anselmi}

\vskip .1truecm

\textit{Dipartimento di Fisica ``Enrico Fermi", Universit\`{a} di Pisa}

\textit{Largo B. Pontecorvo 3, 56127 Pisa, Italy}

\textit{and INFN, Sezione di Pisa,}

\textit{Largo B. Pontecorvo 3, 56127 Pisa, Italy}

damiano.anselmi@unipi.it

\vskip2truecm

\textbf{Abstract}
\end{center}

We formulate quantum field theories of massive fields of arbitrary spins.
The presence of both physical and fake particles, organized into multiplets,
makes it possible to fulfill the requirements of locality, unitarity and
renormalizability at the same time. The theories admit cubic and quartic
self-interactions and can be coupled to quantum gravity and gauge fields.
The simplest irreducible bosonic and fermionic multiplets are made of towers
of alternating physical and fake particles. Their mass spectrum is
constrained by RG invariant relations and depends on just one or two masses.
The fixed points of the renormalization-group flow are scale invariant, but
not necessarily conformal invariant. The Palatini version of quantum gravity
with fakeons is equivalent to the non-Palatini one coupled to a peculiar
multiplet of order 3. As a consequence, it is equally renormalizable and
unitary.

\vfill\eject

\section{Introduction}

\label{intro}\setcounter{equation}{0}

Massive vectors, spin 2 particles and spin 3/2 particles are described by
the Proca, Pauli-Fierz and Rarita-Schwinger theories \cite%
{proca,paulifierz,rarita}, respectively, which are unitary at the free-field
level, but not renormalizable once interactions are turned on. In addition,
the Pauli-Fierz mass term is known to create pathologies \cite{patho} on
nontrivial backgrounds and when nonlinearities are taken into account, due
to the Boulware-Deser ghost \cite{deser}. Proposals to deal with this issue
have been put forward, such as the de Rham-Gabadadze-Tolley model \cite{drgt}%
, the compactification of five-dimensional theories \cite{five} and the
theories of bimetric gravity \cite{bigravity}. However, these approaches do
not address the renormalizability issue, so the problem of embedding the
Proca, Pauli-Fierz and Rarita-Schwinger fields into a renormalizable,
unitary framework has remained essentially open, so far.

In this paper, we offer a solution by means of purely virtual particles \cite%
{wheelerons}. The key concept, inspired by the theory of quantum gravity
formulated in \cite{LWgrav}, is the idea of \textquotedblleft
fakeon\textquotedblright\ \cite{fakeons}, that is to say, a fake particle
that mediates interactions, but does not belong to the spectrum of
asymptotic states. It is introduced by means of a novel quantization
prescription for the poles of the free propagators, alternative to the
Feynman $i\epsilon $ one. Once the fake degrees of freedom are projected
away, they are not resuscitated back by the radiative corrections. This
makes the prescription/projection consistent with unitarity to all orders 
\cite{fakeons}. Quantum gravity is described by a triplet made of the
graviton, a massive scalar and a massive spin 2 fakeon.

The easiest way to make calculations with the fakeon prescription is to
start from the Euclidean framework and perform the Wick rotation in a new
way. Specifically, the Wick rotation is completed as usual (that is to say
analytically by means of the Feynman prescription) in the Euclidean region
and below the production thresholds that do not involve fakeons. Above the
thresholds that involve fakeons it is completed nonanalytically by means of
the average continuation \cite{LWformulation,fakeons}, which is the
arithmetic average of the two analytic continuations (the Feynman one and
its conjugate). A recent account of how this works, with calculations and
comparison with alternative concepts, in particular a suggestion \cite%
{bollini} based on the Feynman-Wheeler absorber-emitter theory \cite{FW},
can be found in ref. \cite{wheelerons}. For the proofs to all orders the
reader is referred to ref. \cite{fakeons}. The good news are that the
calculations of Feynman diagrams with the fakeon prescription in quantum
gravity \cite{UVQG,absograv} are no harder than analogous calculations in
the standard model. Moreover, once the procedure is extended to curved
backgrounds, the theory makes precise predictions on the power spectra of
inflationary cosmology \cite{ABP}, which will be tested as soon as new
cosmological data become available \cite{CMBStage4}.

The fakeon prescription works irrespective of the sign of the residue at the
pole, under the (no-tachyon) condition that the squared mass have a positive
real part. It can be used to eliminate the ghosts at the fundamental level,
but it cannot be used to eliminate tachyons. Physical particles can also be
turned into fake ones.

In this paper, we make sense of massive particles of arbitrary spins by
embedding them into multiplets made of physical and fake particles,
organized in a way that makes the nonrenormalizable behaviors mutually
cancel out. The simplest irreducible bosonic multiplets are described by
traceless, symmetric tensors $\chi _{\mu _{1}\cdots \mu _{s}}$ of order $s$.
They contain physical particles of spins $s$, $s-2$, $\ldots $, alternating
with fakeons of spins $s-1$, $s-3$, $\ldots $. Their mass spectrum is
uniquely determined by the first two masses. The relations among the masses
are renormalization-group (RG) invariant. The simplest fermionic multiplets
are described by completely symmetric spinors $\psi _{\mu _{1}\cdots \mu
_{s}}$ that satisfy $\gamma ^{\rho }\psi _{\rho \mu _{2}\cdots \mu _{s}}=0$.
They have a similar alternating structure made of physical and fake
particles. Their mass spectrum is determined by a unique mass. Multiplets
with more involved structures exist -- the possibilities being numerous --
and nontrivial mixings among multiplets are allowed.

The multiplets admit renormalizable interactions similar to the ones we are
accustomed to. In four dimensions, the bosonic vertices are quartic and
nonderivative, or cubic with at most one derivative. The interactions
between bosons and fermions are of the Yukawa type. The multiplets can be
coupled to quantum gravity and gauge fields in the usual ways. In the paper,
we study a number of examples, compute the beta functions and check the
optical theorem at one loop.

Moreover, we show that the fixed points of the RG flow have unexpected
features. They are scale invariant, but generically not conformal invariant.
With the help of large $N $ expansions we build some examples and address
the construction of others.

As an application, we study the Palatini version of quantum gravity with
fakeons and show that it is nothing but the non-Palatini one coupled to a
peculiar (reducible) multiplet of order 3. For this reason, it is equally
renormalizable and unitary, once the parameters fulfill the no-tachyon
condition.

A recent analysis of the conditions under which the order-3 tensor just
mentioned propagates only physical particles has been done by Percacci and
Sezgin in \cite{percacci}. Due to the absence of fakeons, their models are
not renormalizable when interactions are switched on.

It may be interesting to inquire whether fakeons can be helpful in the
context of interacting higher-spin gauge fields. Equations of motion for
such fields have been proposed by Vasiliev \cite{vasiliev} (but lack a
satisfactory Lagrangian formulation \cite{higherspin}) and a light-front
theory has been developed by Ponomarev and Skvortsov \cite{skvortsov}. At
present we cannot answer in the affirmative, since fake particles should
stay massive, and possibly heavy, in realistic models. The reason is that
they trigger violations of microcausality \cite{causalityQG} at energies
larger than their masses, so a massless fakeon would be responsible for the
violation of causality at all energies. Nevertheless, we do discuss some
properties of the massless limits when we talk about RG fixed points.

Unless stated otherwise, we work in four spacetime dimensions. The
generalization to other dimensions is straightforward.

The paper is organized as follows. In section \ref{bosons} we discuss the
simplest irreducible bosonic multiplets at the free-field level, while
section \ref{fermions} is devoted to the fermionic multiplets. Section \ref%
{examples} contains a number of examples and the discussion of reducible
multiplets. In section \ref{interactions} we investigate the renormalizable
interactions. We study a simple quartic model at one loop, compute its beta
functions and check that the absorptive parts satisfy the optical theorem.
In sections \ref{fixed} we study the fixed points of the RG flow. In section %
\ref{palatini} we study the Palatini version of quantum gravity with fakeons
and in section \ref{palamul} we discuss the Palatini multiplet. Section \ref%
{conclusions} contains the conclusions.

\section{Bosonic multiplets}

\label{bosons}\setcounter{equation}{0}

In this section we study the bosonic multiplets, starting from the order-1
field $\chi _{\mu }$. The unique two-derivative local Lagrangian we can
write for it is%
\begin{equation}
\mathcal{L}_{1}=-\frac{1}{2}\left[ (\partial _{\rho }\chi _{\mu })(\partial
^{\rho }\chi ^{\mu })+a_{1}(\partial ^{\nu }\chi _{\nu })(\partial _{\rho
}\chi ^{\rho })-m_{1}^{2}\chi _{\mu }\chi ^{\mu }\right] ,  \label{l1}
\end{equation}%
up to total derivatives. Defininig the spin 1 projector%
\begin{equation}
\pi _{\mu \nu }=\eta _{\mu \nu }-\frac{p_{\mu }p_{\nu }}{p^{2}},
\label{proj1}
\end{equation}%
the propagator%
\begin{equation}
\langle \chi _{\mu }(p)\chi _{\nu }(-p)\rangle =-\frac{i\pi _{\mu \nu }}{%
p^{2}-m_{1}^{2}+i\epsilon }-\frac{m_{0}^{2}}{m_{1}^{2}}\left. \frac{p_{\mu
}p_{\nu }}{p^{2}}\frac{i}{p^{2}-m_{0}^{2}}\right\vert _{\mathrm{f}}
\label{pro1}
\end{equation}%
has a spin-1 pole of mass $m_{1}$ and a spin-0 pole of squared mass%
\begin{equation}
m_{0}^{2}=\frac{m_{1}^{2}}{1+a_{1}}.  \label{pmn}
\end{equation}

The no-tachyon condition gives $a_{1}>-1$. When it holds, the pole at $%
p^{2}=m_{1}^{2}$ has a residue with the correct sign for a physical
particle, while the pole at $p^{2}=m_{0}^{2}$ has a residue with the wrong
sign. This means that the spin-0 particle must be quantized as a fakeon,
while the spin-1 particle can be quantized as a physical particle or a
fakeon. Throughout this paper we choose to maximize the number of physical
particles, but it is understood that the fakeon prescription can be adopted
to turn any subset of them into fake particles.

If the overall sign of (\ref{l1}) is flipped, the roles of the two poles are
exchanged and we obtain a multiplet made of a fake vector and a physical
scalar. From now on, we choose the overall sign so that the particle of
highest spin is physical, unless stated otherwise. The prescriptions of
formula (\ref{pro1}) conform to the convention just stated, so the spin-1
pole is treated by means of the Feynman $i\epsilon $ prescription, while the
scalar pole is treated by means of the fakeon prescription, denoted by the
subscript \textquotedblleft f\textquotedblright .

The propagator (\ref{pro1}) has the right behavior at large $p^{2}$ to
ensure renormalizability. The projectors (\ref{proj1}) and $p_{\mu }p_{\nu
}/p^{2}$ introduce spurious poles at $p^{2}=0$, which cancel out in the sum.
The prescriptions of (\ref{pro1}) refer to the poles at $p^{2}=m_{1}^{2}$
and $p^{2}=m_{0}^{2}$, while the spurious $p^{2}=0$ poles can be prescribed
the way we like, as long as use the same convention for all of them. For
definiteness, here and below we understand that the $p^{2}=0$ poles are
defined by means of the Feynman prescription $p^{2}\rightarrow
p^{2}+i\epsilon $. It is possible to rewrite the propagator (\ref{pro1}) in
the form%
\begin{equation}
\langle \chi _{\mu }(p)\chi _{\nu }(-p)\rangle =-\frac{i}{%
p^{2}-m_{1}^{2}+i\epsilon }\left( \eta _{\mu \nu }-\frac{p_{\mu }p_{\nu }}{%
m_{1}^{2}}\right) -\left. \frac{p_{\mu }p_{\nu }}{m_{1}^{2}}\frac{i}{%
p^{2}-m_{0}^{2}}\right\vert _{\mathrm{f}},  \label{pro1abs}
\end{equation}%
where its regularity for $p^{2}\rightarrow 0$ becomes evident, but it is no
longer evident that the behavior for large $p^{2}$ is the right one for
renormalizability.

Note that the two-point function of the $\chi _{\mu }$ divergence, which is 
\begin{equation*}
\langle \partial ^{\mu }\chi _{\mu }(p)\hspace{0.01in}\partial ^{\nu }\chi
_{\nu }(-p)\rangle =-\frac{m_{0}^{2}p^{2}}{m_{1}^{2}}\left. \frac{i}{%
p^{2}-m_{0}^{2}}\right\vert _{\mathrm{f}},
\end{equation*}%
gets rid of the vector and highlights the scalar. The Proca theory is
obtained by letting $m_{0}^{2}$ tend to infinity, i.e., studying the limit $%
a_{1}\rightarrow -1^{+}$.

Summarizing, with the conventions stated above the massive vector $\chi
_{\mu }$ describes a multiplet made of a spin-1 physical particle and a
spin-0 fakeon.

The second example we consider is the spin 2 multiplet. Let $\chi _{\mu \nu
} $ denote a symmetric, traceless tensor. Up to total derivatives, its
unique Lagrangian is%
\begin{equation}
\mathcal{L}_{2}=\frac{1}{2}\left[ (\partial _{\rho }\chi _{\mu \nu
})(\partial ^{\rho }\chi ^{\mu \nu })+a_{2}(\partial ^{\nu }\chi _{\mu \nu
})(\partial _{\rho }\chi ^{\rho \mu })-m_{2}^{2}\chi _{\mu \nu }\chi ^{\mu
\nu }\right] .  \label{l2}
\end{equation}%
The propagator%
\begin{equation}
\langle \chi _{\mu \nu }(p)\chi _{\rho \sigma }(-p)\rangle =\frac{%
i\prod\nolimits_{\mu \nu ,\rho \sigma }^{(2)}(p)}{p^{2}-m_{2}^{2}+i\epsilon }%
-\frac{m_{1}^{2}}{2m_{2}^{2}}\left. \frac{i\prod\nolimits_{\mu \nu ,\rho
\sigma }^{(1)}(p)}{p^{2}-m_{1}^{2}}\right\vert _{\mathrm{f}}+\frac{3m_{0}^{2}%
}{4m_{2}^{2}}\frac{i\prod\nolimits_{\mu \nu ,\rho \sigma }^{(0)}(p)}{%
p^{2}-m_{0}^{2}+i\epsilon }  \label{pro2}
\end{equation}%
has a spin-2 pole of mass $m_{2}$, plus a spin-1 pole and a spin-0 pole of
squared masses%
\begin{equation}
m_{1}^{2}=\frac{2m_{2}^{2}}{2+a_{2}},\qquad m_{0}^{2}=\frac{4m_{2}^{2}}{%
4+3a_{2}},  \label{mass2}
\end{equation}%
respectively. The residues at the poles involve the tensors%
\begin{eqnarray*}
\prod\nolimits_{\mu \nu ,\rho \sigma }^{(2)}(p) &=&\frac{1}{2}\left( \pi
_{\mu \rho }\pi _{\nu \sigma }+\pi _{\mu \sigma }\pi _{\nu \rho }-\frac{2}{3}%
\pi _{\mu \nu }\pi _{\rho \sigma }\right) ,\qquad \prod\nolimits_{\mu \nu
,\rho \sigma }^{(0)}(p)=\tilde{\pi}_{\mu \nu }\tilde{\pi}_{\rho \sigma }, \\
\prod\nolimits_{\mu \nu ,\rho \sigma }^{(1)}(p) &=&-\frac{1}{p^{2}}(p_{\mu
}p_{\rho }\pi _{\nu \sigma }+p_{\mu }p_{\sigma }\pi _{\nu \rho }+p_{\nu
}p_{\rho }\pi _{\mu \sigma }+p_{\nu }p_{\sigma }\pi _{\mu \rho }),
\end{eqnarray*}%
where%
\begin{equation*}
\tilde{\pi}_{\mu \nu }=\frac{4}{3}\left( \frac{p_{\mu }p_{\nu }}{p^{2}}-%
\frac{\eta _{\mu \nu }}{4}\right) .
\end{equation*}

Again, the propagator (\ref{pro2}) has no pole for $p^{2}=0$. The fakeon
prescription in (\ref{pro2}) just refers to the $p^{2}=m_{1}^{2}$ pole,
while it is understood that the spurious poles at $p^{2}=0$ are defined by
means of the Feynman prescription. It is possible to rewrite (\ref{pro2}) as%
\begin{equation}
\langle \chi _{\mu \nu }(p)\chi _{\rho \sigma }(-p)\rangle =\frac{i\left.
\prod\nolimits_{\mu \nu ,\rho \sigma }^{(2)}(p)\right\vert
_{p^{2}\rightarrow m_{2}^{2}}}{p^{2}-m_{2}^{2}+i\epsilon }-\frac{m_{1}^{2}}{%
2m_{2}^{2}}\left. \frac{i\left. \prod\nolimits_{\mu \nu ,\rho \sigma
}^{(1)}(p)\right\vert _{p^{2}\rightarrow m_{1}^{2}}}{p^{2}-m_{1}^{2}}%
\right\vert _{\mathrm{f}}+\frac{3m_{0}^{2}}{4m_{2}^{2}}\frac{i\left.
\prod\nolimits_{\mu \nu ,\rho \sigma }^{(0)}(p)\right\vert
_{p^{2}\rightarrow m_{0}^{2}}}{p^{2}-m_{0}^{2}+i\epsilon },  \label{pro2abso}
\end{equation}%
where the $p^{2}=0$ poles are manifestly gone, but renormalizability is no
longer evident. It is understood that the replacements $p^{2}\rightarrow
m_{i}^{2}$ of (\ref{pro2abso}) are formal, i.e. they act only on the
denominators, but do not constrain the components $p^{\mu }$.

From (\ref{mass2}) we read the no-tachyon condition, which is $a_{2}>-4/3$.
The three masses of the multiplet are related to one another, because they
depend on just two parameters. Their relation,%
\begin{equation}
\frac{1}{m_{2}^{2}}-\frac{3}{m_{1}^{2}}+\frac{2}{m_{0}^{2}}=0,  \label{mass}
\end{equation}%
is renormalization-group invariant, so when (renormalizable) interactions
are switched on (see section \ref{interactions}) formula (\ref{mass})
continues to hold with the masses replaced by the running ones. RG invariant
relations among parameters are usually due to nonanomalous symmetries. The
remarkable feature of relations like (\ref{mass}) is that they are mere
consequences of power counting.

We can highlight the particles of lower spins from the two-point functions
of multiple divergences: 
\begin{eqnarray*}
\mathrm{Res}[\langle \partial ^{\nu }\chi _{\mu \nu }(p)\hspace{0.01in}%
\partial ^{\sigma }\chi _{\rho \sigma }(-p)\rangle ]_{p^{2}=m_{1}^{2}} &=&%
\frac{im_{1}^{4}}{2m_{2}^{2}}\left. \pi _{\mu \rho }\right\vert
_{p^{2}=m_{1}^{2}}, \\
\mathrm{Res}[\langle \partial ^{\mu }\partial ^{\nu }\chi _{\mu \nu }(p)%
\hspace{0.01in}\partial ^{\rho }\partial ^{\sigma }\chi _{\rho \sigma
}(-p)\rangle ]_{p^{2}=m_{0}^{2}} &=&\frac{3im_{0}^{6}}{4m_{2}^{2}}.
\end{eqnarray*}%
With the overall sign of (\ref{l2}) the spin-1 particle is a fakeon, while
the spin-2 and spin-0 particles can be physical. The prescriptions of (\ref%
{pro2}) were chosen in anticipation of this.

We cannot obtain the Pauli-Fierz theory as a limit of this case, since we
are using a traceless tensor $\chi _{\mu \nu }$. We will be able to obtain
it in section \ref{examples}.

Summarizing, the order-2 massive field $\chi _{\mu \nu }$ describes a
multiplet made of a spin-2 particle, a spin-1 fakeon and a spin-0 particle.

\subsection{Arbitrary spin}

Now we move to the case of arbitrary spin $s$. Let $\chi _{\mu _{1}\cdots
\mu _{s}}$ denote a traceless, completely symmetric tensor of order $s$. The
most general two-derivative quadratic Lagrangian we can build for $\chi
_{\mu _{1}\cdots \mu _{s}}$ in flat space is%
\begin{equation}
\mathcal{L}_{s}=\frac{(-1)^{s}}{2}\left[ (\partial _{\mu }\chi _{\mu
_{1}\cdots \mu _{s}})(\partial ^{\mu }\chi ^{\mu _{1}\cdots \mu
_{s}})+a_{s}(\partial ^{\mu }\chi _{\mu \mu _{2}\cdots \mu _{s}})(\partial
_{\nu }\chi ^{\nu \mu _{2}\cdots \mu _{s}})-m_{s}^{2}\chi _{\mu _{1}\cdots
\mu _{s}}\chi ^{\mu _{1}\cdots \mu _{s}}\right] ,  \label{ls}
\end{equation}%
up to total derivatives. We find the propagator%
\begin{equation}
\langle \chi _{\mu _{1}\cdots \mu _{s}}(p)\hspace{0.01in}\chi _{\nu
_{1}\cdots \nu _{s}}(-p)\rangle =\sum_{n=0}^{s}\left. \frac{i(-1)^{s-n}r_{sn}%
}{p^{2}-m_{sn}^{2}}\right\vert _{\sigma _{sn}}{\prod }_{\mu _{1}\cdots \mu
_{s},\nu _{1}\cdots \nu _{s}}^{(s,n)}(p),  \label{propa}
\end{equation}%
with the mass spectrum%
\begin{equation}
m_{sn}^{2}=m_{s}^{2}\lambda _{sn},\qquad \lambda _{sn}=\left[ 1+\frac{a_{s}}{%
2s^{2}}(s-n)(s+n+1)\right] ^{-1},  \label{lsn}
\end{equation}%
and the coefficients%
\begin{equation}
r_{sn}=\lambda _{sn}\frac{n!(s-n)!(s+n+1)!}{2^{s}(s!)^{2}(2n+1)!!}.
\label{res}
\end{equation}%
The subscript $\sigma _{sn}$ denotes the quantization prescription.
Anticipating the results we are going to find, $\chi _{\mu _{1}\cdots \mu
_{s}}$ describes a multiplet made of alternating physical particles and
fakeons of spins $s$, $s-1$, \ldots\ $0$. Specifically, $\sigma _{sn}$
stands for the Feynman $i\epsilon $ prescription for $n=s$, $s-2$, etc.,
while it stands for the fakeon prescription for $n=s-1$, $s-3$, etc.

The tensor ${\prod }_{\mu _{1}\cdots \mu _{s},\nu _{1}\cdots \nu
_{s}}^{(s,n)}(p)$ is dimensionless, built only with $\eta _{\mu \nu }$ and $%
p_{\mu }$, and uniquely determined by the following properties:

($i$) it is completely symmetric and traceless in each $s$-tuple $\mu
_{1}\cdots \mu _{s}$ and $\nu _{1}\cdots \nu _{s}$;

($ii$) it is symmetric under the exchange of the two $s$-tuples;

($iii$) it is \textquotedblleft $(s-n+1)$-transverse\textquotedblright ,
i.e. it vanishes if contracted with $p_{\mu _{1}}\cdots p_{\mu _{s-n+1}}$;

($iv$) it contains at most $n$ tensors $\eta _{\mu \nu }$ with indices of
both $s$-tuples;

($v$) finally, it is normalized so that 
\begin{equation}
{\prod }_{i_{1}\cdots i_{n}0\cdots 0,j_{1}\cdots j_{n}0\cdots 0}^{(s,n)}(p)=%
\mathbb{I}_{i_{1}\cdots i_{n},j_{1}\cdots j_{n}}^{(3)}  \label{space}
\end{equation}%
in the rest frame $p^{\mu }=(p^{0},0,0,0)$, where $i_{1}\cdots i_{n}$ and $%
j_{1}\cdots j_{n}$ are space indices and $\mathbb{I}_{i_{1}\cdots
i_{n},j_{1}\cdots j_{n}}^{(3)}$ is the identity operator for order-$n$
symmetric, traceless tensors in three space dimensions, built with the
Kronecker delta $\delta _{ij}$.

Let us see how to build these tensors more explicitly. If we take $\mathbb{I}%
_{i_{1}\cdots i_{s},j_{1}\cdots j_{s}}^{(3)}$ and replace every tensor $%
\delta _{ij}$ with $\pi _{\mu \nu }$ (paying attention to the relabeling of
the indices), the result, which is clearly traceless and transverse,
coincides with $(-1)^{s}{\prod }_{\mu _{1}\cdots \mu _{s},\nu _{1}\cdots \nu
_{s}}^{(s,s)}(p)$, so it satisfies (\ref{space}). From it, we can build the
tensor%
\begin{equation}
{\prod }_{\mu _{1}\cdots \mu _{s},\nu _{1}\cdots \nu _{s}}^{(s,s-1)}=\frac{%
s^{2}}{p^{2}}p_{\{\mu _{s}}{\prod }_{\mu _{1}\cdots \mu _{s-1}\},\{\nu
_{1}\cdots \nu _{s-1}}^{(s-1,s-1)}p_{\nu _{s}\}},  \label{ps1}
\end{equation}%
which is manifestly $2$-transverse and traceless in each $s$-tuple (complete
symmetrization over indices between curly brackets being understood).
Moreover, it is normalized so as to satisfy (\ref{space}). Next, we have%
\begin{equation}
{\prod }_{\mu _{1}\cdots \mu _{s},\nu _{1}\cdots \nu _{s}}^{(s,s-2)}=\frac{%
s^{2}(s-1)^{2}}{4}\tilde{\pi}_{\{\mu _{s-1}\mu _{s}}{\prod }_{\mu _{1}\cdots
\mu _{s-2}\},\{\nu _{1}\cdots \nu _{s-2}}^{(s-2,s-2)}\tilde{\pi}_{\nu
_{s-1}\nu _{s}\}},  \label{ps2}
\end{equation}%
where now%
\begin{equation}
\tilde{\pi}_{\mu \nu }=\frac{2s}{2s-1}\left( \frac{p_{\mu }p_{\nu }}{p^{2}}-%
\frac{\eta _{\mu \nu }}{2s}\right) .  \label{ptmn}
\end{equation}%
Again, (\ref{ps2}) is traceless and $3$-transverse, and satisfies (\ref%
{space}). The other tensors can be worked out similarly. Specifically, we
find 
\begin{equation}
{\prod }_{\mu _{1}\cdots \mu _{s},\nu _{1}\cdots \nu _{s}}^{(s,n)}=\binom{s}{%
n}^{2}\tilde{\pi}_{\{\mu _{n+1}\cdots \mu _{s}}{\prod }_{\mu _{1}\cdots \mu
_{n}\},\{\nu _{1}\cdots \nu _{n}}^{(n,n)}\tilde{\pi}_{\nu _{n+1}\cdots \nu
_{s}\}},  \label{ps3}
\end{equation}%
where $\tilde{\pi}_{\mu _{n+1}\cdots \mu _{s}}$ is a tensor built with $\eta
_{\mu \nu }$ and $p_{\mu }$, completely symmetric and normalized so that $%
\tilde{\pi}_{0\cdots 0}=1$. It is proportional to $p_{\mu _{n+1}}\cdots
p_{\mu _{s}}$ plus terms obtained by replacing pairs of momenta $p_{\mu
}p_{\nu }$ with $\eta _{\mu \nu }p^{2}$, multiplied by relative coefficients
arranged to make (\ref{ps3}) traceless in the each $s$-tuple.

The particle of spin $n$ can be isolated from the rest by taking $s-n$
divergences and the appropriate residue:%
\begin{eqnarray}
&&\mathrm{Res}[\langle \partial ^{\mu _{s-n+1}}\cdots \partial ^{\mu
_{s}}\chi _{\mu _{1}\cdots \mu _{s}}(p)\hspace{0.01in}\partial ^{\nu
_{s-n+1}}\cdots \partial ^{\nu _{s}}\chi _{\nu _{1}\cdots \nu
_{s}}(-p)\rangle ]_{p^{2}=m_{sn}^{2}}=  \notag \\
&&\qquad \qquad \qquad \qquad \qquad =\left. i(-1)^{s-n}r_{sn}m_{sn}^{2(s-n)}%
{\prod }_{\mu _{1}\cdots \mu _{n},\nu _{1}\cdots \nu
_{n}}^{(n,n)}(p)\right\vert _{p^{2}=m_{sn}^{2}}.  \label{diva}
\end{eqnarray}

A way to determine the coefficients of (\ref{propa}) is to require that the
singularities for $p^{2}\rightarrow 0$ (with $p^{\mu }$ nonvanishing)
mutually cancel. Indeed, the Lagrangian (\ref{ls}) implies that when the
squared mass $m_{s}^{2}$ is large we must have%
\begin{equation*}
\langle \chi _{\mu _{1}\cdots \mu _{s}}(p)\hspace{0.01in}\chi _{\nu
_{1}\cdots \nu _{s}}(-p)\rangle \sim \frac{i(-1)^{s+1}}{m_{s}^{2}}\mathbb{I}%
_{\mu _{1}\cdots \mu _{s}\nu _{1}\cdots \nu _{s}},
\end{equation*}%
where $\mathbb{I}_{\mu _{1}\cdots \mu _{s}\nu _{1}\cdots \nu _{s}}$ is the
identity operator for order-$s$ symmetric traceless tensors (in four
dimensions), built with $\eta _{\mu \nu }$. Moreover, the expansion in
powers of $p^{\mu }/m_{s}$ is regular.

To make sense of $\mathcal{L}_{s}$ as a theory of physical particles and
fakeons, we must require that the propagator (\ref{propa}) contain no
tachyonic poles, i.e., $\lambda _{sn}>0$ for every $n=0,\ldots ,s$. This
means%
\begin{equation}
a_{s}>-\frac{2s}{s+1}\equiv \bar{a}_{s},  \label{as}
\end{equation}%
which we will assume henceforth. If $a_{s}<0$ ($a_{s}>0$) the particle of
highest spin is also the least (most) massive one and the masses increase
(decrease) when the spin increases.

The factor $(-1)^{s-n}$ that appears in (\ref{propa}) and (\ref{diva}) shows
that the multiplet is made of alternating physical particles and fakeons of
spins $s$, $s-1$, etc. Precisely, the particles of spins $s$, $s-2$, $s-4$,
etc., can be quantized as physical particles by means of the Feynman $%
i\epsilon $ prescription, while the particles of spins $s-1$, $s-3$, $s-5$,
etc., must be quantized as fakeons.

Formula (\ref{propa}) shows that the propagator falls off as $1/p^{2}$ for
large $|p^{2}|$ and generic values of $a_{s}$, which is the right behavior
for renormalizability. The locality of counterterms can be proved in the
usual fashion (starting in Euclidean space and performing the Wick rotation
either analytically or by means of the average continuation \cite{fakeons}).
The terms behaving in a nonrenormalizable way, typical of the Proca and
Pauli-Fierz propagators, mutually cancel among the particles of the
multiplet.

The RG invariant relations among the squared masses $m_{sn}^{2}=m_{s}^{2}%
\lambda _{sn}$ are%
\begin{equation}
\frac{(s-n-1)(s+n)}{m_{s}^{2}}-\frac{(s-n)(s+n+1)}{m_{s,s-1}^{2}}+\frac{2s}{%
m_{sn}^{2}}=0,\qquad n=0,\ldots ,s.  \label{RGmas}
\end{equation}

An alternative form of the propagator (\ref{propa}) is obtained by formally
replacing $p^{2}$ with $m_{sn}^{2}$ inside each ${\prod }_{\mu _{1}\cdots
\mu _{s},\nu _{1}\cdots \nu _{s}}^{(s,n)}(p)$. Then the regularity for $%
p^{2}\rightarrow 0$ becomes apparent, but it is no longer evident that the
behavior for large momenta is the right one for renormalizability.

With the help of Mathematica, we have studied the multiplets of orders 3, 4
and 5 and explicitly verified that their propagators agree with (\ref{propa}%
) and subsequent formulas.

\section{Fermionic multiplets}

\label{fermions}\setcounter{equation}{0}

In this section we study the simplest fermionic multiplets. Consider a
spinor $\psi _{\mu _{1}\cdots \mu _{s}}$ that is completely symmetric in $%
\mu _{1}\cdots \mu _{s}$ and satisfies%
\begin{equation}
\gamma ^{\rho }\psi _{\rho \mu _{2}\cdots \mu _{s}}=0.  \label{gam}
\end{equation}%
Due to this condition, $\psi _{\mu _{1}\cdots \mu _{s}}$ is automatically
traceless in its spacetime indices, as can be proved by contracting it with
two $\gamma $ matrices.

The unique one-derivative local Lagrangian we can build for $\psi _{\mu
_{1}\cdots \mu _{s}}$ is%
\begin{equation}
\mathcal{L}_{s+1/2}=(-1)^{s}\bar{\psi}_{\mu _{1}\cdots \mu _{s}}\left(
i\gamma ^{\rho }\partial _{\rho }-m_{s}\right) \psi ^{\mu _{1}\cdots \mu
_{s}},  \label{lsf}
\end{equation}%
up to total derivatives, since every other term vanishes due to (\ref{gam}).

It is useful to introduce the projector for the condition (\ref{gam}) \cite%
{HSCF}. Assuming that $\Psi ^{\mu _{1}\cdots \mu _{s}}$ does not satisfy (%
\ref{gam}), then%
\begin{equation*}
\tilde{\Psi}^{\mu _{1}\cdots \mu _{s}}\equiv \Psi ^{\mu _{1}\cdots \mu _{s}}-%
\frac{1}{2(s+1)}\sum_{i=1}^{s}\gamma ^{\mu _{i}}\gamma _{\alpha }\Psi ^{\mu
_{1}\cdots \mu _{i-1}\alpha \mu _{i+1}\cdots \mu _{s}}
\end{equation*}%
does. With the help of this, we can easily derive the field equations of (%
\ref{lsf}), which read%
\begin{equation*}
i\gamma ^{\rho }\partial _{\rho }\psi ^{\mu _{1}\cdots \mu _{s}}-\frac{i}{s+1%
}\sum_{i=1}^{s}\gamma ^{\mu _{i}}\partial _{\alpha }\psi ^{\mu _{1}\cdots
\mu _{i-1}\alpha \mu _{i+1}\cdots \mu _{s}}-m_{s}\psi ^{\mu _{1}\cdots \mu
_{s}}=0.
\end{equation*}%
Contracting with $\partial _{\mu _{1}}\cdots \partial _{\mu _{s-n}}$, we
find the RG invariant mass spectrum%
\begin{equation}
m_{sn}=\frac{s+1}{n+1}m_{s},\qquad n=0,1,\ldots s.  \label{RGmasf}
\end{equation}

The propagators exhibit features similar to the ones of the bosonic
multiplets of the previous section. In particular, the residues at the poles
with $n=$ odd have the same relative signs, which are opposite to the signs
of the residues at the poles with $n=$ even. With the overall sign of (\ref%
{lsf}), the poles with $n=s$, $s-2$, etc., can be quantized as physical
particles by means of the Feynman prescription, while the poles with $n=s-1$%
, $s-3$, etc., must be quantized as fakeons. The roles of the two subsets of
poles are exchanged if the overall sign of (\ref{lsf}) is flipped.

In the limit of vanishing masses we get the higher-spin conformal fermionic
fields of ref. \cite{HSCF}. In section \ref{fixed} we discuss scale
invariant and conformal field theories in detail.

As an example, consider the spin 3/2 multiplet. It propagates a massive
particle of spin 3/2 and a massive fakeon of spin 1/2. The Lagrangian%
\begin{equation}
\mathcal{L}_{3/2}=-\bar{\psi}_{\mu }\left( i\gamma ^{\rho }\partial _{\rho
}-m\right) \psi ^{\mu }  \label{l32}
\end{equation}%
gives the propagator%
\begin{eqnarray}
\langle \psi _{\mu }(p)\hspace{0.01in}\bar{\psi}_{\nu }(-p)\rangle &=&-\frac{%
i}{p^{2}-m^{2}+i\epsilon }P_{\mu \rho }\left[ (\gamma ^{\alpha }p_{\alpha
}+m)\eta ^{\rho \sigma }-\frac{2}{3}\frac{p^{\rho }p^{\sigma }}{p^{2}}%
(\gamma ^{\alpha }p_{\alpha }+2m)\right] P_{\sigma \nu }  \notag \\
&&-\frac{8}{3}\left. \frac{i}{p^{2}-4m^{2}}\right\vert _{\mathrm{f}}P_{\mu
\rho }\frac{p^{\rho }p^{\sigma }}{p^{2}}(\gamma ^{\alpha }p_{\alpha
}+2m)P_{\sigma \nu },  \label{pr32}
\end{eqnarray}%
where 
\begin{equation*}
P_{\mu \nu }=\eta _{\mu \nu }-\frac{1}{4}\gamma _{\mu }\gamma _{\nu }
\end{equation*}%
is the \textquotedblleft projector\textquotedblright\ for the condition (\ref%
{gam}) [$P_{\mu \rho }P_{%
\phantom \rho%
\nu }^{\rho }=P_{\mu \nu }$, $(P_{\mu \nu })^{\dagger }=\gamma ^{0}P_{\mu
\nu }\gamma ^{0}$]. As always, the $p^{2}=0$ poles of (\ref{pr32}) are
spurious. The spin 1/2 particle can be isolated by evaluating the two-point
function of the divergence:%
\begin{equation}
\langle \partial ^{\mu }\psi _{\mu }(p)\hspace{0.01in}\partial ^{\nu }\bar{%
\psi}_{\nu }(-p)\rangle =-\frac{3ip^{2}}{2}\left. \frac{\gamma ^{\alpha
}p_{\alpha }+2m}{p^{2}-4m^{2}}\right\vert _{\mathrm{f}}.  \label{dive}
\end{equation}%
We see that the sign of its residue is opposite to the one of a common Dirac
particle.

Note that it is not possible to obtain the Rarita-Schwinger theory as a
limit of this model, since the masses of the two particles are related,
which prevents us from sending the fakeon mass to infinity while keeping the
other mass finite. We will describe how to retrieve the Rarita-Schwinger
theory in the next section. Letting aside the fakeon prescription for the
spin 1/2 particle, a model related to (\ref{l32}) was considered by
Haberzettl in \cite{Haberzettl} in the context of nuclear resonances.

We have also checked the spin 5/2 multiplet, which is made of a physical
spin 5/2 particle of mass $m$, a spin 3/2 fakeon of mass $3m/2$ and a
physical spin 1/2 particle of mass $3m$.

\section{Further examples}

\label{examples}\setcounter{equation}{0}

In this section we discuss other examples, starting from the antisymmetric
tensor $A_{\mu \nu }$, which describes a particle and a fakeon, both of spin
1. The Lagrangian is%
\begin{equation}
\mathcal{L}_{A}=\frac{1}{2}(\partial _{\mu }A_{\nu \rho })(\partial ^{\mu
}A^{\nu \rho })+\frac{b}{2}(\partial ^{\mu }A_{\mu \rho })(\partial _{\nu
}A^{\nu \rho })-\frac{m^{2}}{2}A_{\mu \nu }A^{\mu \nu },  \label{la}
\end{equation}%
with propagator%
\begin{equation}
\langle A_{\mu \nu }(p)\hspace{0.01in}A_{\rho \sigma }(-p)\rangle =\frac{%
i(\pi _{\mu \rho }\pi _{\nu \sigma }-\pi _{\mu \sigma }\pi _{\nu \rho })}{%
2(p^{2}-m^{2}+i\epsilon )}+\left. \frac{i(p_{\mu }p_{\rho }\eta _{\nu \sigma
}-p_{\mu }p_{\sigma }\eta _{\nu \rho }-p_{\nu }p_{\rho }\eta _{\mu \sigma
}+p_{\nu }p_{\sigma }\eta _{\mu \rho })}{(2+b)p^{2}(p^{2}-\bar{m}^{2})}%
\right\vert _{\mathrm{f}},  \label{proa}
\end{equation}%
where%
\begin{equation*}
\bar{m}^{2}=\frac{2m^{2}}{2+b}.
\end{equation*}%
The no-tachyon condition is $b>-2$, which implies that the two poles have
residues of opposite signs, so one of them is a fake degree of freedom. In (%
\ref{proa}) the pole of mass $m$ is a physical particle, while the other
pole describes a fakeon.

Completely antisymmetric tensors with three and four indices, $A_{\mu \nu
\rho }$ and $A_{\mu \nu \rho \sigma }$, can be converted to a vector and a
scalar, respectively, by means of Hodge dualization, that is to say, by
contracting them with $\varepsilon ^{\mu \nu \rho \sigma }$.

We can also study bosonic tensors and fermionic tensors with assorted
symmetry properties. The variety is huge, but at least one of them, the most
general bosonic tensor $\Omega _{\mu \nu \rho }$ of order 3, is worth of
interest, since it allows us to describe the Palatini version of quantum
gravity with fakeons. For this reason, we postpone its analysis to section %
\ref{palamul}.

Particles with the same spin belonging to different irreducible multiplets
can mix in nontrivial ways. For example, a traceful symmetric tensor $\tilde{%
\chi}_{\mu \nu }$ gives a reducible multiplet containing a spin 2 particle,
a spin 1 fakeon and two mixing scalar particles. Splitting $\tilde{\chi}%
_{\mu \nu }$ into its traceless part $\chi _{\mu \nu }$ and its trace $\chi $%
, the Lagrangian is%
\begin{eqnarray}
\mathcal{\tilde{L}}_{2} &=&\frac{1}{2}\left[ (\partial _{\rho }\chi _{\mu
\nu })(\partial ^{\rho }\chi ^{\mu \nu })\pm (\partial _{\mu }\chi
)(\partial ^{\mu }\chi )+a_{2}(\partial ^{\nu }\chi _{\mu \nu })(\partial
_{\rho }\chi ^{\rho \mu })\right.  \notag \\
&&\qquad \qquad +\left. a_{2}^{\prime }(\partial ^{\nu }\chi _{\mu \nu
})(\partial ^{\mu }\chi )-m^{2}(\chi _{\mu \nu }\chi ^{\mu \nu
}+a_{2}^{\prime \prime }\chi ^{2})\right] .  \label{l2p}
\end{eqnarray}%
We have fixed the overall sign to keep the spin 2 particle physical. The
term $(\partial _{\mu }\chi )(\partial ^{\mu }\chi )$ is normalized apart
from its sign.

Working out the propagator, it is possible to check that the spin 2 particle
is physical and has squared mass $m^{2}$, while the spin 1 particle is fake
and has squared mass $2m^{2}/(2+a_{2})$, the no-tachyon condition being $%
a_{2}>-2$. The scalar poles solve an equation of the form $P_{2}(p^{2})=0$,
where $P_{2}$ is a polynomial of degree two. The solutions $m_{i}^{2}$, $%
i=1,2$, may be real or complex. The no-tachyon condition becomes the
requirement that $m_{i}^{2}$ have positive real parts. If the squared masses 
$m_{i}^{2}$ are complex, they describe the so-called thick fakeons, which
are fakeons with nontrivial widths at the fundamental level (which means: in
the free-field limit, before including radiative corrections such as the
self energies). The thick fakeons are also the fake particles that appear in
the (reformulated) Lee-Wick models (see \cite{LWformulation,causalityQG}).
If the squared masses $m_{i}^{2}$ are real, we may have two physical
scalars, two fake scalars or one and one, depending on the region in the
space of parameters $a_{2}$, $a_{2}^{\prime }$, $a_{2}^{\prime \prime }$.

We can obtain the Pauli-Fierz action (with an unusual normalization of the
trace) as a limit of (\ref{l2p}) for $a_{2}\rightarrow -2$, $a_{2}^{\prime
}\rightarrow 2\sqrt{2/3}$, $a_{2}^{\prime \prime }\rightarrow -2$, choosing
the minus sign in front of $(\partial _{\mu }\chi )(\partial ^{\mu }\chi )$.
The spin 1 fakeon and the scalar particles become infinitely massive in such
a limit. Moreover, the behavior of the propagator for large $|p^{2}|$
becomes incompatible with renormalizability.

If the spin 2 particle were not embedded in the multiplet $\tilde{ \chi}%
_{\mu \nu }$, a mass term different from the Pauli-Fierz one ($a_{2}^{\prime
\prime }\neq -2$) would create problems. The reason why such problems do not
show up here is that the slot of the would-be Boulware-Deser ghost is
already occupied by a fake degree of freedom and properly treated by means
of the fakeon prescription.

Examples of mixing kinetic terms for multiplets of different spins are%
\begin{equation*}
\bar{\psi}_{\mu \mu _{2}\cdots \mu _{s}}\partial ^{\mu }\psi ^{\mu
_{2}\cdots \mu _{s}},\qquad \chi _{\mu \mu _{2}\cdots \mu _{s}}^{\prime
}\partial ^{\mu }\chi ^{\mu _{2}\cdots \mu _{s}},\qquad \chi _{\mu \nu \mu
_{3}\cdots \mu _{s}}^{\prime }\partial ^{\mu }\partial ^{\nu }\chi ^{\mu
_{3}\cdots \mu _{s}}.
\end{equation*}

The Rarita-Schwinger theory can be obtained as follows. Consider the spin
3/2 multiplet of (\ref{l32}) and mix it with a Dirac spin 1/2 particle $\psi 
$. This gives a theory that describes one spin 3/2 particle and
(generically) two spin 1/2 fakeons. Enough new parameters\ are introduced
(through the coefficients of $\bar{\psi}\psi $, $\bar{\psi}\partial ^{\mu
}\psi _{\mu }$ and $\bar{\psi}_{\mu }\partial ^{\mu }\psi $) to untie the
masses. At that point, we can let the masses of both fakeons tend to
infinity and keep the mass of the spin 3/2 particle finite at the same time.
Again, renormalizability is lost in the limit.

\section{Interacting theories}

\label{interactions}\setcounter{equation}{0}

In this section we study the interactions, under the constraints of
locality, renormalizability and unitarity. Since the quadratic parts of the
Lagrangians met so far have two derivatives in the case of bosons and one in
the case of fermions, the dimensions of the fields in units of mass are the
usual ones (1 for bosons, 3/2 for fermions). The vertices may contain: four
bosons without derivatives; three bosons with one derivative at most; two
fermions and one boson without derivatives. We check renormalizability and
unitarity in a relatively simple model by computing the renormalization
constants and the absorptive parts at one loop.

\subsection[chimunu]{The $\chi _{\mu \nu }^{4}$ theory}

The most general interacting theory of a traceless order-2 multiplet $\chi
_{\mu \nu }$ is described by the Lagrangian%
\begin{equation}
\mathcal{L}_{2}^{\text{int}}=\mathcal{L}_{2}-\frac{\lambda }{4!}\chi _{\mu
\nu }\chi ^{\nu \rho }\chi _{\rho \sigma }\chi ^{\sigma \mu }-\frac{\lambda
^{\prime }}{4!}(\chi _{\mu \nu }\chi ^{\mu \nu })^{2},  \label{l4}
\end{equation}%
where $\mathcal{L}_{2}$ is given in (\ref{l2}) and $\lambda $, $\lambda
^{\prime }$ are the coupling constants. The model can be renormalized by
means of a wave-function renormalization constant ($\chi _{\mu \nu
}\rightarrow Z_{\chi }^{1/2}\chi _{\mu \nu }$), a mass renormalization ($%
m_{2}^{2}\rightarrow m_{2}^{2}Z_{m}$), plus renormalizations of the
parameters ($\lambda \rightarrow \lambda +\delta \lambda $, $\lambda
^{\prime }\rightarrow \lambda ^{\prime }+\delta \lambda ^{\prime }$, $%
a_{2}\rightarrow a_{2}+\delta a_{2}$).

Using the propagator (\ref{pro2}), it is straightforward to compute these
quantities at one loop. As in the scalar $\varphi ^{4}$ theory, $Z_{\chi }$
is uncorrected and so is $a_{2}$. Instead, $Z_{m}$ is corrected by the
tadpole diagram. We find%
\begin{equation*}
Z_{m}=1+\frac{15a_{2}^{4}+100a_{2}^{3}+288a_{2}^{2}+384a_{2}+192}{576\pi
^{2}\varepsilon \left( 3a_{2}^{2}+10a_{2}+8\right) {}^{2}}\left( 19\lambda
+44\lambda ^{\prime }\right) ,
\end{equation*}%
where $\varepsilon =4-D$ and $D$ is the continued dimension of the
dimensional regularization. In passing, this result confirms that the
relation (\ref{mass}) among the masses of the three particles that belong to
the multiplet is RG invariant, since all of them have the same
renormalization constant.

We see that the dependence on $a_{2}$ is rather nontrivial, the reason being
that we are treating this parameter exactly. Although we have computed $%
\delta \lambda $ and $\delta \lambda ^{\prime }$ for generic $a_{2}$, we
report the results only to the first order in $a_{2}$ around $a_{2}=0$: 
\begin{equation*}
\delta \lambda =\frac{\lambda (\lambda +3\lambda ^{\prime })}{12\pi
^{2}\varepsilon }\left( 1-\frac{a_{2}}{2}\right) +\mathcal{O}%
(a_{2}^{2}),\qquad \delta \lambda ^{\prime }=\frac{33\lambda ^{2}+152\lambda
\lambda ^{\prime }+272\lambda ^{\prime 2}}{768\pi ^{2}\varepsilon }\left( 1-%
\frac{a_{2}}{2}\right) +\mathcal{O}(a_{2}^{2}).
\end{equation*}%
The one-loop beta functions are $\beta _{\lambda }=\varepsilon \delta
\lambda $, $\beta _{\lambda ^{\prime }}=\varepsilon \delta \lambda ^{\prime
} $, $\beta _{a_{2}}=0$.

Unitarity can be studied by evaluating the absorptive parts of the diagrams.
We follow the technique developed in refs. \cite{UVQG,absograv}, which
allows us to disentangle the contributions of the physical particles from
those of the fakeons in an efficient way. For this purpose, it is convenient
to reorganize the propagator in the form (\ref{pro2abso}), to eliminate the
spurious poles at $p^{2}=0$.

The bubble diagram%
\begin{equation}
\raisebox{-1mm}{\scalebox{2}{$\rangle\!{\bigcirc\!{\langle}}$}}  \label{bub}
\end{equation}%
can be decomposed as the sum of nine contributions, which we denote by $%
B_{jk}$, where the indices $j,k=2,1,0$ refer to the poles of the propagators
of the internal legs, decomposed as in (\ref{pro2abso}). Each $B_{jk}$ has
branch cuts above the threshold $p^{2}=(m_{j}+m_{k})^{2}$, where $p^{\mu }$
denotes the external momentum.

The calculation is divided in three steps. First, we switch to the Euclidean
framework, where we make use of Feynman parameters and integrate on the loop
momentum. Second, we perform the Wick rotation to Minkowski spacetime,
concentrating on the region located below the thresholds, which is the
region where the external momentum $p^{\mu }$ satisfies $p^{2}\leqslant
(m_{j}+m_{k})^{2}$ for every $j$ and $k$. Once we subtract the divergent
part and take the limit $D\rightarrow 4$, we find%
\begin{equation}
B_{jk}=-i\int_{0}^{1}\mathrm{d}x\hspace{0.01in}\left\{ Q_{jk}(p,x)\ln \left[
m_{j}^{2}(1-x)+m_{k}^{2}x-p^{2}x(1-x)\right] +Q_{jk}^{\prime }(p,x)\right\} ,
\label{bij}
\end{equation}%
where $Q_{jk}(p,x)$ and $Q_{jk}^{\prime }(p,x)$ are certain real polynomials
of $x$ and $p^{\mu }$.

Third, we overcome the thresholds as follows. In the cases $B_{22}$, $B_{20}$%
, $B_{02}$ and $B_{00}$, where no fakeon is involved, we complete the Wick
rotation analytically by means of the Feynman prescription. The result we
obtain is (\ref{bij}) with the replacement $p^{2}\rightarrow p^{2}+i\epsilon 
$. The absorptive part is%
\begin{equation}
2\mathrm{Im}[(-i)B_{jk}]=2\pi \int_{0}^{1}\mathrm{d}x\hspace{0.01in}%
Q_{jk}(p,x)\theta (p^{2}x(1-x)-m_{j}^{2}(1-x)-m_{k}^{2}x).  \label{abso}
\end{equation}

In the cases $B_{21}$, $B_{12}$, $B_{11}$, $B_{10}$ and $B_{01}$, where one
or two circulating fakeons are present, we complete the Wick rotation
nonanalytically by means of the average continuation, which is the
arithmetic average of the two analytic continuations $p^{2}\rightarrow
p^{2}\pm i\epsilon $. As a consequence, the absorptive parts vanish.

The optical theorem \cite{veltman,diagrammar} can be expressed
diagrammatically by means of the identity

\begin{equation}
2\hspace{0.01in}\mathrm{Im}\left[ (-i)\raisebox{-1mm}{\scalebox{2}{$\rangle%
\!{\bigcirc\!{\langle}}$}}\right] =\raisebox{-1mm}{\scalebox{2}{$\rangle\!{%
\bigcirc\!{\langle}}\hspace{-0.16in}\slash\hspace{0.1in}$}}=\sum_{f}\int 
\mathrm{d}\Pi _{f}\hspace{0.01in}\left\vert \raisebox{-1mm}{\scalebox{2}{$%
\rangle\!\!{\hspace{0.02in}<}$}}\right\vert ^{2},  \label{cutd}
\end{equation}%
where the sum is over the physical final states $f$, while the integral is
performed on their phase space $\Pi _{f}$. The states $f$ involved here are:
($i$) a pair of spin 2 particles; ($ii$) a pair of spin 0 particles; and ($%
iii$) a pair made of one spin 2 particle and one spin 0 particle. Excluded
final states are all those that involve the spin 1 fakeon.

The cut diagram appearing in the middle of equation (\ref{cutd}) actually
stands for the sum of two cut diagrams, depending on which side of the cut
is \textquotedblleft shadowed\textquotedblright . Define the cut propagators%
\begin{equation}
P_{\mu \nu \rho \sigma }^{\pm }(p)=2\pi \theta (\pm p^{0})\left[ \left.
\prod\nolimits_{\mu \nu ,\rho \sigma }^{(2)}(p)\right\vert
_{p^{2}\rightarrow m_{2}^{2}}\delta (p^{2}-m_{2}^{2})+\frac{3m_{0}^{2}}{%
4m_{2}^{2}}\left. \prod\nolimits_{\mu \nu ,\rho \sigma }^{(0)}(p)\right\vert
_{p^{2}\rightarrow m_{0}^{2}}\delta (p^{2}-m_{0}^{2})\right] .
\label{cutpro}
\end{equation}%
Note that the fakeon poles do not contribute to the right-hand side. The two
cut diagrams of (\ref{cutd}) can be evaluated by integrating the products 
\begin{equation*}
P_{\mu \nu \rho \sigma }^{+}(k)P_{\alpha \beta \gamma \delta
}^{+}(p-k),\qquad P_{\mu \nu \rho \sigma }^{-}(k)P_{\alpha \beta \gamma
\delta }^{-}(p-k),
\end{equation*}%
respectively [where $k^{\mu }$ and $(p-k)^{\mu }$ are the momenta of the
internal legs]. The results are multiplied by the vertices, the
combinatorial factor 1/2 and a factor $(-1)$ for the shadowed vertex.

Now, the left-hand side of (\ref{cutd}) receives only the contributions (\ref%
{abso}) for $jk=22$, $20$, $02$ and $00$, but so does the right-hand side
and the cut diagrams in the middle, as emphasized by (\ref{cutpro}).
Moreover, each $B_{jk}$ can be thought of as a diagram in its own. For $%
jk=22 $, $20$, $02$ and $00$ such a diagram only involves the Feynman
prescription, so it satisfies the identity (\ref{cutd}) by the usual
(\textquotedblleft pre-fakeon\textquotedblright ) arguments \cite%
{veltman,diagrammar,unitarityc}. Recalling that each $B_{jk}$ with $jk=21$, $%
12$, $11$, $10$ and $01$ gives zero contribution to the left-hand side of (%
\ref{cutd}), we conclude that the bubble diagram (\ref{bub}) of the $\chi
_{\mu \nu }^{4}$ theory does satisfy the optical theorem in the subspace of
physical states, in agreement with unitarity.

We see that making computations with arbitrary-spin massive multiplets does
not require more effort than making computations in the standard model and
quantum gravity with fakeons \cite{UVQG,absograv}. Moreover, the level of
difficulty of the fakeon prescription is comparable to the one of the
Feynman prescription.

\subsection{Other interactions}

A simple model that couples the system (\ref{l4}) to the spin 3/2 multiplet
is the one described by the Lagrangian%
\begin{equation}
\mathcal{L}_{Y}=\mathcal{L}_{2}^{\text{int}}+\mathcal{L}_{3/2}+\lambda _{Y}%
\bar{\psi}_{\mu }\chi ^{\mu \nu }\psi _{\nu },  \label{inter}
\end{equation}%
where $\mathcal{L}_{3/2}$ is given in formula (\ref{l32}) and $\lambda _{Y}$
is a Yukawa coupling.

Examples of vertices we can build for more general multiplets are%
\begin{eqnarray}
&&\bar{\psi}_{\mu _{1}\cdots \mu _{s}}\chi ^{\mu _{1}\cdots \mu _{s}\nu
_{1}\cdots \nu _{r}}\psi _{\nu _{1}\cdots \nu _{r}},\qquad \bar{\psi}_{\mu
_{1}\cdots \mu _{s}}\gamma _{\alpha }\chi ^{\alpha \mu _{1}\cdots \mu
_{s}\nu _{1}\cdots \nu _{r}}\psi _{\nu _{1}\cdots \nu _{r}},  \notag \\
&&\bar{\psi}_{\mu _{1}\cdots \mu _{s}}\sigma _{\alpha \beta }A^{\alpha \beta
\mu _{1}\cdots \mu _{s}\nu _{1}\cdots \nu _{r}}\psi _{\nu _{1}\cdots \nu
_{r}},\qquad \chi _{\mu \mu _{2}\cdots \mu _{s}}^{(1)}(\partial ^{\mu }\chi
_{\nu _{1}\cdots \nu _{t}}^{(2)\mu _{2}\cdots \mu _{r}})\chi ^{(3)\mu
_{r+1}\cdots \mu _{s}\nu _{1}\cdots \nu _{t}},  \label{more}
\end{eqnarray}%
where $\sigma _{\alpha \beta }=-i[\gamma _{\alpha },\gamma _{\beta }]/2$ and 
$A^{\alpha \beta \mu _{1}\cdots \mu _{s}\nu _{1}\cdots \nu _{r}}$ is
antisymmetric in $\alpha \beta $. Of course, we can also have versions with $%
\gamma _{5}$ and $\varepsilon ^{\mu \nu \rho \sigma }$.

The coupling to gravity is built with the help of the tetrad formalism,
where the conditions of vanishing traces and the condition (\ref{gam}) can
be expressed algebraically, without involving the metric tensor $g_{\mu \nu
} $. We have interacting theories of completely symmetric traceless bosons $%
\chi _{a_{1}\cdots a_{s}}$, where $a_{1}\cdots a_{s}$ are indices of the
tangent space, completely symmetric fermions $\psi _{a_{1}\cdots a_{t}}$
satisfying the condition $\gamma ^{a}\psi _{aa_{2}\cdots a_{t}}=0$, mixing
multiplets and multiplets with more complicated structures. The actions need
to be covariantized and equipped with all the renormalizable nonminimal
couplings we can build.

It is possible to couple the multiplets to QED, non-Abelian gauge fields and
the whole standard model in the usual fashion. For example, the Lagrangian
of $SU(N)$ Yang-Mills theories coupled to the fermionic models of (\ref{lsf}%
) is 
\begin{equation}
\mathcal{L}_{\text{YM}}=-\frac{1}{4}F_{\mu \nu }^{a}F^{a\mu \nu }+(-1)^{s}%
\bar{\psi}_{\mu _{1}\cdots \mu _{s}}(i\gamma ^{\rho }\partial _{\rho
}-ig\gamma ^{\rho }A_{\rho }^{a}T^{a}-m_{s})\psi ^{\mu _{1}\cdots \mu _{s}},
\label{qed}
\end{equation}%
where $g$ is the gauge coupling, $a$ is the index of the adjoint
reprentation and $T^{a}$ are the anti-Hermitian matrices of the matter
representation.

\section{Fixed points of the RG flow}

\label{fixed}\setcounter{equation}{0}

In this section we study the fixed points of the RG\ flow in quantum field
theories of interacting multiplets of arbitrary spins. We show that they are
in general scale invariant, but not conformal invariant, with some
exceptions.

The RG flow, the renormalization constants, the beta functions and the
anomalous dimensions are the same for any combination of Feynman/fakeon
prescriptions we may choose for the poles of the free propagators \cite%
{LWgrav,fakeons}. The reason is that, due to the locality of counterterms,
the overall divergent parts of the diagrams have a trivial Wick rotation.
This implies that they can be computed below every threshold or, which is
the same, in Euclidean space, where no prescription is necessary.

For the moment, we concentrate on the Euclidean theories, which give
information on a subsector of the Minkowskian ones. For example, the
Euclidean fixed points allow us to calculate critical exponents and the
central charges $a$ and $c$ \cite{HSCF}, as well as establish whether the
models satisfy the $a$ theorem \cite{freed} or not. Later we discuss the
Wick rotation to Minkowski spacetime. We keep using the Minkowskian notation
throughout the discussion.

We start by noting that if we set 
\begin{equation}
m_{s}=0,\qquad a_{s}=\bar{a}_{s}=-\frac{2s}{s+1},  \label{confo}
\end{equation}%
in (\ref{ls}), we obtain the higher-spin conformal fields introduced in ref. 
\cite{HSCF}. Up to total derivatives, the $\chi _{\mu _{1}\cdots \mu _{s}}$
Lagrangian can be written in this case as%
\begin{equation}
\hspace{0.01in}\mathcal{L}_{s}^{\text{C}}=\frac{(-1)^{s}}{4}\hspace{0.01in}%
F_{\mu \nu \mu _{2}\cdots \mu _{s}}F^{\mu \nu \mu _{2}\cdots \mu _{s}},
\label{lsc}
\end{equation}%
where%
\begin{equation}
F_{\mu \nu \mu _{2}\cdots \mu _{s}}=\partial _{\mu }\chi _{\nu \mu
_{2}\cdots \mu _{s}}-\partial _{\nu }\chi _{\mu \mu _{2}\cdots \mu _{s}}-%
\frac{1}{s+1}\sum_{i=2}^{s}(\eta _{\mu \mu _{i}}\partial ^{\alpha }\chi
_{\alpha \nu \mu _{2}\cdots \hat{\mu}_{i}\cdots \mu _{s}}-\eta _{\nu \mu
_{i}}\partial ^{\alpha }\chi _{\alpha \mu \mu _{2}\cdots \hat{\mu}_{i}\cdots
\mu _{s}})  \label{fs}
\end{equation}%
is a sort of field strength (the hat denoting omitted indices), invariant
under the gauge symmetry 
\begin{equation}
\delta \chi _{\mu _{1}\cdots \mu _{s}}=\partial _{\mu _{1}}\cdots \partial
_{\mu _{s}}\Lambda -\mathrm{traces.}  \label{gau}
\end{equation}

The proof that the action 
\begin{equation}
S_{s}^{\text{C}}=\int \mathrm{d}^{4}x\hspace{0.01in}\mathcal{L}_{s}^{\text{C}%
}  \label{sc}
\end{equation}%
is conformal invariant follows from the property that the full conformal
group can be reconstructed by composing the Poincar\'{e} group with the
coordinate inversion $x^{\mu }\rightarrow x^{\mu }/x^{2}$, under which the
field strength transforms as%
\begin{equation*}
F_{\mu \nu \mu _{2}\cdots \mu _{s}}\rightarrow (x^{2})^{2}I_{\mu \rho
}I_{\nu \sigma }I_{\mu _{2}\rho _{2}}\cdots I_{\mu _{s}\rho _{s}}F^{\rho
\sigma \rho _{2}\cdots \rho _{s}},
\end{equation*}%
where $I_{\mu \nu }=\eta _{\mu \nu }-2(x_{\mu }x_{\nu }/x^{2})$. Clearly, (%
\ref{sc}) is invariant under the coordinate inversion. Since it is obviously
invariant under the Poincar\'{e} group, it is conformal invariant.

If we insert $a_{s}=\bar{a}_{s}$ into (\ref{lsn}) we obtain 
\begin{equation}
\lambda _{sn}=\frac{s(s+1)}{n(n+1)}.  \label{n0}
\end{equation}%
Letting $a_{s}\rightarrow \bar{a}_{s}$ in (\ref{ls}), the scalar belonging
to the multiplet, which has squared mass $m_{s}^{2}\lambda _{s0}$, becomes
infinitely heavy and drops out of the spectrum. Then the propagator (\ref%
{propa}) becomes nonrenormalizable (of the
Proca/Pauli-Fierz/Rarita-Schwinger type). The reason is that the missing
scalar must make room for the gauge symmetry (\ref{gau}) when $%
m_{s}\rightarrow 0$.

The field strength\ (\ref{fs}) allows us to uncover interesting properties
also when the conditions (\ref{confo}) do not hold and the symmetry (\ref%
{gau}) is absent. Adding total derivatives, we can recast the Lagrangian (%
\ref{ls}) in the equivalent form%
\begin{equation}
\mathcal{L}_{s}^{\prime }=\frac{(-1)^{s}}{2}\hspace{0.01in}\left[ \frac{1}{2}%
F_{\mu \nu \mu _{2}\cdots \mu _{s}}F^{\mu \nu \mu _{2}\cdots \mu
_{s}}+(a_{s}-\bar{a}_{s})(\partial ^{\mu }\chi _{\mu \mu _{2}\cdots \mu
_{s}})(\partial _{\nu }\chi ^{\nu \mu _{2}\cdots \mu _{s}})-m_{s}^{2}\chi
_{\mu _{1}\cdots \mu _{s}}\chi ^{\mu _{1}\cdots \mu _{s}}\right] .
\label{ls2}
\end{equation}%
The remarkable feature of this expression is that, due to (\ref{as}), $-%
\mathcal{L}_{s}^{\prime }$ is positive definite in the Euclidean framework.

The positive definiteness can be extended to interacting theories. For
example, if we assume $\lambda >0$, $\lambda ^{\prime }>0$, the Lagrangian 
\begin{equation}
\mathcal{L}_{2\hspace{0.01in}\mathrm{int}}(a_{2},m_{2},\lambda ,\lambda
^{\prime })=\mathcal{L}_{2}^{\prime }-\frac{\lambda }{4!}\chi _{\mu \nu
}\chi ^{\nu \rho }\chi _{\rho \sigma }\chi ^{\sigma \mu }-\frac{\lambda
^{\prime }}{4!}(\chi _{\mu \nu }\chi ^{\mu \nu })^{2}  \label{l2int}
\end{equation}%
is also negative-definite in Euclidean space. More generally, we can
consider the interacting theories%
\begin{eqnarray}
\mathcal{L}_{s\hspace{0.01in}\mathrm{int}}(a_{s},m_{s},\lambda ) &=&\frac{%
(-1)^{s}}{4}\hspace{0.01in}F_{\mu \nu \mu _{2}\cdots \mu _{s}}F^{\mu \nu \mu
_{2}\cdots \mu _{s}}  \notag \\
&&+\frac{(-1)^{s}}{2}(a_{s}-\bar{a}_{s})(\partial ^{\mu }\chi _{\mu \mu
_{2}\cdots \mu _{s}})(\partial _{\nu }\chi ^{\nu \mu _{2}\cdots \mu
_{s}})-V(\chi ,m_{s},\lambda ),  \label{lsint}
\end{eqnarray}%
where the potential $V$ collects the mass term of (\ref{ls2}) plus quartic
interactions, $\lambda $ denoting the couplings. As long as $V$ is positive
definite in Euclidean space, so is $-\mathcal{L}_{s\hspace{0.01in}\mathrm{int%
}}$.

The fixed points of the RG flow are the solutions of the conditions of
vanishing beta functions. Such conditions ensure scale invariance, but not
necessarily conformal invariance. To explain this point, let us refer to (%
\ref{lsint}) for concreteness. Since the running of the masses is
proportional to the masses themselves, it is consistent to switch the masses
off at all energies and focus on massless theories. Recalling that we are
still concentrating on the Euclidean versions of the theories, the massless
limits present no particular difficulties. The classical action%
\begin{equation}
S_{s\hspace{0.01in}\mathrm{int}}(a_{s},0,\lambda )=\int \mathrm{d}^{4}x%
\hspace{0.01in}\mathcal{L}_{s\hspace{0.01in}\mathrm{int}}(a_{s},0,\lambda )
\label{sint}
\end{equation}%
is obviously scale invariant. The quantum field theory derived from $S_{s%
\hspace{0.01in}\mathrm{int}}$ is not scale invariant, in general, because
the couplings may run. It becomes scale invariant for the values $\lambda
^{\ast }$, $a_{s}^{\ast }$ of $\lambda $ and $a_{s}$ where the beta
functions vanish: 
\begin{equation}
\beta _{\lambda }(\lambda ^{\ast },a_{s}^{\ast })=\beta _{a_{s}}(\lambda
^{\ast },a_{s}^{\ast })=0.  \label{betas}
\end{equation}%
When these conditions are satisfied, we can construct a conserved charge
associated with the dilatation.

However, a fixed point that satisfies (\ref{betas}) is not necessarily
conformal invariant. Consider the free-field limit (\ref{ls}). If we set $%
m_{s}=0$ and keep $a_{s}\neq \bar{a}_{s}$, the theory is scale invariant and
the beta functions are trivially zero. However, it is not conformal
invariant: it becomes so only for $a_{s}\rightarrow \bar{a}_{s}$.

Going back to the general case, a necessary condition to have conformal
invariance is that the classical action (\ref{sint}) be conformal, which
happens if and only if%
\begin{equation}
a_{s}^{\ast }=\bar{a}_{s}.  \label{ase}
\end{equation}%
However, it is difficult to fulfill (\ref{ase}) and (\ref{betas}) at the
same time, because in total they make a system containing more equations
than unknowns. It may be possible to solve the system when extra parameters
are present, if they are nonrunning (possibly because protected by
nonanomalous symmetries).

Other difficulties come from the gauge invariance (\ref{gau}), since it is
hard to build interacting theories that preserve it. Because of these
problems, the existence of nontrivial conformal field theories involving
bosonic multiplets $\chi _{\mu _{1}\cdots \mu _{s}}$ remains a conjecture
for the moment. We content ourselves with the existence of scale invariant
fixed points.

Similar conclusions apply to the other types of bosonic multiplets, but for
different reasons. For example, let us consider the antisymmetric tensor $%
A_{\mu \nu }$, with Lagrangian (\ref{la}). We know that the no-tachyon
condition requires $b>-2$. When $b$ tends to $-2$ and $m$ tends to zero, the
Lagrangian (\ref{la}) becomes gauge invariant, the gauge symmetry being $%
\delta A_{\mu \nu }=\partial _{\mu }C_{\nu }-\partial _{\nu }C_{\mu }$. It
becomes conformal invariant only when $b=-4$, $m=0$ \cite{HSCF}. Values of $%
b $ as low as $-4$ are deep into the tachyonic region for $m\neq 0$, so it
might be difficult to reach them from the quantum field theories we are
studying here. On the contrary, the RG flow can easily reach a scale
invariant fixed point without $b$ leaving the no-tachyon region.

The fermionic multiplets $\psi _{\mu _{1}\cdots \mu _{s}}$ of section \ref%
{fermions} have nicer properties. Indeed, their free Lagrangian (\ref{lsf})
does not contain independent parameters other than the mass. In the massless
limit, the action%
\begin{equation}
S_{s+1/2}^{C}=\int \mathrm{d}^{4}x\hspace{0.01in}\mathcal{L}%
_{s+1/2}^{C}=\int \mathrm{d}^{4}x\hspace{0.01in}\bar{\psi}_{\mu _{1}\cdots
\mu _{s}}i\gamma ^{\rho }\partial _{\rho }\psi ^{\mu _{1}\cdots \mu _{s}}
\label{lsfc}
\end{equation}%
is invariant under the full conformal group (see \cite{HSCF} for details).

Let us consider interacting theories of multiplets $\psi _{\mu _{1}\cdots
\mu _{s}}$, scalar fields, fermions of spin 1/2 and gauge fields in the
massless limit. We exclude bosonic multiplets, as well as gravity. An
example of such theories is the one with Lagrangian 
\begin{equation}
\mathcal{L}_{\text{YM}}^{\prime }=-\frac{1}{4}F_{\mu \nu }^{a}F^{a\mu \nu }+%
\bar{\psi}(i\gamma ^{\rho }\partial _{\rho }-ig\gamma ^{\rho }A_{\rho }^{a}%
\bar{T}^{a})\psi +(-1)^{s}\bar{\psi}_{\mu _{1}\cdots \mu _{s}}(i\gamma
^{\rho }\partial _{\rho }-ig\gamma ^{\rho }A_{\rho }^{a}T^{a})\psi ^{\mu
_{1}\cdots \mu _{s}},  \label{lim}
\end{equation}%
where $s>0$ and $\bar{T}^{a}$ are the matrices of another representation of
the gauge group. The classical actions are conformal invariant. We can show
that (the Euclidean versions of) the fixed points are conformal invariant.

The easiest way to prove this fact is by working in coordinate space.
Vertices and propagators transform as they should under the conformal group,
because they transform correctly under the Poincar\'{e} group and the\
coordinate inversion. Knowing this, it is easy to prove that each Feynman
diagram transforms correctly, as long as it is convergent. On the other
hand, the overall divergent parts are local and have the same structures as
the terms of the classical action, which are conformal invariant. The
violation of conformal invariance comes only from the scale introduced by
the regularization and the trace anomaly associated with it. Specifically,
the conserved, symmetric energy-momentum tensor has a trace that does not
vanish on shell. However, at the fixed points of the RG flow, where the beta
functions are zero, the trace anomaly does vanish, so the energy-momentum
tensor becomes conserved, symmetric and traceless on shell. This allows us
to build conserved charges for the generators of the whole conformal group
and proves that the fixed point is conformal invariant in the Euclidean
framework.

The Minkowskian versions of these theories deserve a separate discussion.
When fakeons are absent, only the Feynman prescription, which is conformal
invariant, is in play, so the theories remain conformal after the Wick
rotation to Minkowski spacetime. This is nothing new, however, since
dropping the fakeons means dropping the massive multiplets of spins $>1/2$
altogether, so models such as (\ref{lim}) end up containing just gauge
fields, scalars and spin 1/2 fermions. When fakeons are present, the Wick
rotation must be completed nonanalytically \cite{LWformulation,fakeons},
which makes the whole issue of conformal invariance less trivial. Indeed,
the fakeon prescription is formulated in momentum space, where it is
meaningful to talk about $S$ matrix, asymptotic states, optical theorem,
production thresholds and so on. A formulation of the fakeon prescription
directly in coordinate space is still unavailable. Symmetrically, it is not
simple to map the coordinate inversion to momentum space and prove conformal
invariance from the Feynman rules and the Feynman diagrams by working
directly in momentum space. This means that at present we cannot guarantee
that the Minkowskian fixed points are conformal, even when their Euclidean
versions are.

\subsection[Large N limit]{Large $N$ limit}

Other remarkable classes of interacting RG flows and fixed points can be
built with the help of large $N$ expansions. In two dimensions nonderivative
vertices with four higher-spin fermionic legs are renormalizable by power
counting, \`{a} la Gross-Neveu \cite{grossneveu}. In three dimensions they
are renormalizable in the large $N$ expansion \cite{parisi}. In turn, the
three-dimensional models can be used to define interacting conformal field
theories \cite{mio} and build peculiar RG flows interpolating between them,
which can be solved nonperturbatively in the couplings for every truncation
of the large $N$ expansion \cite{mio2}.

Consider the model with Lagrangian 
\begin{equation}
\mathcal{L}^{(N)}=(-1)^{s}\sum_{I=1}^{N}\bar{\psi}_{\mu _{1}\cdots \mu
_{s}}^{I}\left( i\gamma ^{\rho }\partial _{\rho }+\frac{\lambda }{\sqrt{N}}%
\sigma -m\right) \psi _{I}^{\mu _{1}\cdots \mu _{s}}-\frac{M\sigma ^{2}}{2},
\label{largeN}
\end{equation}%
in three spacetime dimensions, where $\psi _{I}^{\mu _{1}\cdots \mu _{s}}$
are four-component spinors, $\sigma $ is an auxiliary field for the four
fermion vertex, $\lambda $ is a coupling that stays finite in the large $N$
limit and $M$, $m$ are masses. The $\sigma $ self energy is leading in the
large $N$ expansion and provides a propagator for $\sigma $. It is possible
to show that the Lagrangians (\ref{largeN}) are renormalizable, once the
perturbative expansion is reorganized as a large $N$ expansion \cite{parisi}.

The massless limit $M,m\rightarrow 0$ of (\ref{largeN}) is smooth in the
Euclidean framework, where the theory becomes conformal invariant, at the
classical level as well as at the quantum level. The conformal limit is
described by the (Euclidean version of the) Lagrangian%
\begin{equation}
\mathcal{L}_{\sigma }^{(N)}=(-1)^{s}\sum_{I=1}^{N}\bar{\psi}_{\mu _{1}\cdots
\mu _{s}}^{I}\left( i\gamma ^{\rho }\partial _{\rho }+\frac{\lambda }{\sqrt{N%
}}\sigma \right) \psi _{I}^{\mu _{1}\cdots \mu _{s}}.  \label{lnsigma}
\end{equation}

For example, in the case of the spin 3/2 multiplet ($s=1$), the $\sigma $
propagator, provided by the leading order of the $\sigma $ two-point
function, turns out to be, in momentum space, 
\begin{equation}
\langle \sigma (p)\hspace{0.01in}\sigma (-p)\rangle =-\frac{2i}{9\lambda ^{2}%
\sqrt{-p^{2}}}+\mathcal{O}\left( \frac{1}{N}\right)  \label{ss}
\end{equation}%
for $p^{2}<0$, which is the Euclidean region. Writing the bare fields as $%
\psi _{\mathrm{B}}^{\mu }=Z_{\psi }^{1/2}\psi ^{\mu }$, $\sigma _{\mathrm{B}%
}=Z_{\sigma }^{1/2}\sigma $, the renormalization constants are%
\begin{equation}
Z_{\psi }=1-\frac{1}{15\pi ^{2}\varepsilon N}+\mathcal{O}\left( \frac{1}{%
N^{2}}\right) ,\qquad Z_{\sigma }=1+\frac{56}{45\pi ^{2}\varepsilon N}+%
\mathcal{O}\left( \frac{1}{N^{2}}\right) ,  \label{resa}
\end{equation}%
where $\varepsilon =3-D$, while the $\lambda $ beta function vanishes
identically. As before, the nonanalytic Wick rotation to Minkowski spacetime
may not preserve conformal invariance in the massless limit, so we limit our
discussion to the Euclidean version for the time being.

It is possible to build simple quantum field theories that interpolate
between pairs of fixed points of the type just met. The prototype of the
models we have in mind has the Lagrangian%
\begin{equation}
\mathcal{L}_{\sigma }^{(N_{1},N_{2})}=(-1)^{s}\sum_{I=1}^{N_{1}}\bar{\psi}%
_{\mu _{1}\cdots \mu _{s}}^{I}\left( i\gamma ^{\rho }\partial _{\rho
}+\lambda \sigma \right) \psi _{I}^{\mu _{1}\cdots \mu
_{s}}+(-1)^{s}\sum_{I=1}^{N_{2}}\bar{\chi}_{\mu _{1}\cdots \mu
_{s}}^{I}\left( i\gamma ^{\rho }\partial _{\rho }+g\sigma \right) \chi
_{I}^{\mu _{1}\cdots \mu _{s}},  \label{interpo}
\end{equation}%
where $\lambda $ and $g$ are couplings. It is clear that for $\lambda =0$
the theory (\ref{interpo}) becomes $\mathcal{L}_{\sigma }^{(N_{2})}$ plus $%
N_{1}$ free multiplets. Instead, for $\lambda =g$ it becomes $\mathcal{L}%
_{\sigma }^{(N_{1}+N_{2})}$. Finally, for $g=0$ it becomes $\mathcal{L}%
_{\sigma }^{(N_{1})}$ plus $N_{2}$ free multiplets. Remarkable features are
that: ($i$) the RG flows are both asymptotically safe and infrared safe; ($%
ii $) the beta function can be worked out exactly in $\alpha \equiv \lambda
^{2}/g^{2}$ at every order of the large $N$ expansion; and ($iii$) there is
an obvious duality $\alpha \leftrightarrow 1/\alpha $, $N_{1}\leftrightarrow
N_{2}$.

The relevant Feynman diagrams are simple adaptations of those of the theory (%
\ref{lnsigma}). In the case of the spin 3/2 multiplet, we find, from (\ref%
{resa}), 
\begin{equation}
\beta _{\alpha }=-\frac{56}{45\pi ^{2}N}\frac{\alpha (1-\alpha )}{1+r\alpha
^{2}}+\mathcal{O}\left( \frac{1}{N^{2}}\right) ,  \label{beta}
\end{equation}%
where $r=N_{1}/N_{2}$ is kept finite in the large $N$ expansion and $N\simeq
N_{1}\simeq N_{2}$. The three fixed points mentioned above are $\alpha =0$, $%
1$, $\infty $, respectively. More details on how to make calculations like
the ones that lead to (\ref{resa}) and (\ref{beta}) can be found in refs. 
\cite{mio,mio2}, together with other properties of these classes of models.

Four dimensional analogues of these RG flows and fixed points can be built
in $SU(N_{c})$ gauge theories coupled to $N_{f}$ fermionic multiplets of
arbitrary spin [check (\ref{qed}) and (\ref{lim})]. For definiteness, we can
imagine that the fermions are in the fundamental representation. A Veneziano
expansion for large $N_{c}$ and large $N_{f}$, with $N_{f}/N_{c}$ fixed \cite%
{veneziano}, makes the corrections to the beta functions negligible from
three loops onwards, while the two-loop beta functions have a form similar
to (\ref{beta}) and admit fixed points \`{a} la Banks-Zaks \cite{bankszaks}.

\subsection{Remarks}

Before concluding this section, some remarks are in order. In realistic
models the fakeons are required to be massive. Indeed, they trigger
violations of microcausality \cite{causalityQG} at energies larger than
their masses, so a massless fakeon would be responsible for the violation of
causality at all energies. The masses make a higher-spin multiplet disappear
at low energies. On the other hand, masses become negligible at high
energies, which justifies the interest for the massless limits. More
generally, we can consider the models obtained by switching off all the
dimensionful parameters (which are the masses, the coefficients of one-leg
vertices and the couplings of nonderivative bosonic 3-leg vertices, if
present). This choice is compatible with the RG equations, since the beta
functions of the dimensionful parameters vanish when those parameters
vanish. The RG\ flows of massless models typically interpolate between
nontrivial IR\ and UV fixed points, as in the examples provided above.

The massless limits and the massless models can be studied straightforwardly
in Euclidean space, as we have already noted. They are more challenging in
Minkowski spacetime. Consider the propagators (\ref{pro2}) and (\ref%
{pro2abso}) for the spin 2 multiplet $\chi _{\mu \nu }$. All the poles
coalesce to $p^{2}=0$ for $m_{2}\rightarrow 0$. Expression (\ref{pro2abso}),
which is free of spurious poles at $p^{2}=0$ and neatly separates the
particles of the multiplet according to their quantization prescriptions, is
not very useful here, because each term has a singular massless limit, since
the residues at the poles contain masses in the denominators. If we use
expression (\ref{pro2}), instead, each term ends up containing products of $%
p^{2}=0$ poles defined by the Feynman prescription, multiplied by $p^{2}=0$
poles defined by the fakeon prescription. Although this does not appear to
pose consistency problems, it is necessary to pay attention to how the limit
is reached.

We conclude by stressing again that the fixed points of the RG flows of
quantum field theories containing massive multiplets of arbitrary spins are
generically scale invariant, but not necessarily conformal invariant.
Nevertheless, in several models the fixed points are conformal invariant in
Euclidean space.

\section{Palatini quantum gravity with fakeons}

\label{palatini}\setcounter{equation}{0}

In this section we show that the Palatini version of quantum gravity with
fakeons is equivalent to the non-Palatini version coupled to a massive
multiplet of a peculiar type, which we call Palatini multiplet.

Let us first discuss the issue in Einstein gravity. The Palatini variation 
\cite{palatini} amounts to treating the connection $\Gamma _{\mu \nu }^{\rho
}$ and the metric tensor $g_{\mu \nu }$ as independent fields. If we apply
it to the action%
\begin{equation}
S(g,\Gamma )=-\frac{M_{\mathrm{Pl}}^{2}}{16\pi }\int \mathrm{d}^{4}x\sqrt{-g}%
R(g,\Gamma ),  \label{pala}
\end{equation}
the field equation $\delta S/\delta \Gamma =0$ equates $\Gamma $ to the
Levi-Civita connection $\Gamma (g)$ \cite{einstein}. Moreover, the Hilbert
action $S_{\mathrm{H}}(g)=S(g,\Gamma (g))$ is obtained once the solution is
inserted back into $S(g,\Gamma )$.

In view of the generalization to quantum gravity, it is convenient to shift $%
\Gamma _{\mu \nu }^{\rho }$ by the Levi-Civita connection. The difference%
\begin{equation}
\Omega _{\mu \nu }^{\rho }\equiv \Gamma _{\mu \nu }^{\rho }-\Gamma _{\mu \nu
}^{\rho }(g)  \label{diffa}
\end{equation}%
transforms as a tensor under diffeomorphisms. We are not assuming that the
connection is metric compatible, nor that the torsion vanishes, so $\Omega
_{\mu \nu }^{\rho }$ must be considered as the most general order-3 tensor,
which we call Palatini tensor. The action $S(g,\Gamma )$ can be rephrased as
a functional of the metric and $\Omega _{\mu \nu }^{\rho }$:%
\begin{equation}
S_{\mathrm{P}}(g,\Omega )=S_{\mathrm{H}}(g)+\frac{M_{\mathrm{Pl}}^{2}}{16\pi 
}\int \mathrm{d}^{4}x\sqrt{-g}g^{\mu \nu }(\Omega _{\mu \sigma }^{\rho
}\Omega _{\rho \nu }^{\sigma }-\Omega _{\mu \nu }^{\rho }\Omega _{\sigma
\rho }^{\sigma }).  \label{sp}
\end{equation}%
The field equations for $\Omega _{\mu \nu }^{\rho }$ give%
\begin{equation}
\Omega _{\mu \nu }^{\rho }=\delta _{\nu }^{\rho }\Omega _{\mu }^{3},\qquad
\Omega _{\mu }^{1}=4\Omega _{\mu }^{3},\qquad \Omega _{\mu }^{2}=\Omega
_{\mu }^{3},  \label{eqpala}
\end{equation}%
where $\Omega _{\mu }^{1}=\Omega _{\mu \nu }^{\nu }$, $\Omega _{\mu
}^{2}=\Omega _{\nu \mu }^{\nu }$ and $\Omega _{\mu }^{3}=g_{\mu \nu }g^{\rho
\sigma }\Omega _{\rho \sigma }^{\nu }$. Once the solution is inserted back
into (\ref{sp}), the last term vanishes.

Note that the equations (\ref{eqpala}) do not imply $\Omega _{\mu \nu
}^{\rho }\equiv 0$, but leave the vector $\Omega _{\mu }^{3}$ free. They
imply $\Omega _{\mu \nu }^{\rho }=0$ under the assumption of metric
compatibility ($\Omega _{\mu \nu }^{\rho }=-g^{\rho \alpha }g_{\nu \beta
}\Omega _{\mu \alpha }^{\beta }$), as well as under the assumption of
vanishing torsion ($\Omega _{\mu \nu }^{\rho }=\Omega _{\nu \mu }^{\rho }$),
but not in the most general case \cite{perc2}.

The difference $S_{\mathrm{P}}(g,\Omega )-S_{\mathrm{H}}(g)$ is not the most
general quadratic, nonderivative term we can build for $\Omega _{\mu \nu
}^{\rho }$. Generically, we can consider 
\begin{equation}
S_{\mathrm{P}}^{\prime }(g,\Omega )=S_{\mathrm{H}}(g)+S_{M}(g,\Omega ),
\label{spp}
\end{equation}%
where 
\begin{equation}
S_{M}(g,\Omega )\equiv \frac{M_{\mathrm{Pl}}^{2}}{16\pi }\int \mathrm{d}^{4}x%
\sqrt{-g}\left[ \Omega _{\mu \nu \rho }\sum_{i=1}^{6}M_{i}\Omega ^{\{\mu \nu
\rho \}_{i}}+\sum_{i,j=1}^{3}\Omega _{\mu }^{i}M_{ij}\Omega ^{j\mu }\right] .
\label{sM}
\end{equation}%
Here $\{\mu \nu \rho \}_{i}$ denotes the $i$th permutation of $\mu \nu \rho $%
, $M_{i}$ and $M_{ij}$ are constants and $\Omega _{\mu \nu \rho }=\Omega
_{\mu \nu }^{\sigma }g_{\rho \sigma }$. There are no compelling reasons to
exclude the additional quadratic terms contained in $S_{M}$. Besides,
renormalization turns them on anyway. Once we switch to $\Gamma _{\mu \nu
}^{\rho }$ by inverting (\ref{diffa}), formula (\ref{spp}) provides a
\textquotedblleft Palatini version of Einstein gravity\textquotedblright\ as
legitimate as (\ref{pala}). The $\Omega _{\mu \nu }^{\rho }$ field equations
of (\ref{spp}) give $\Omega _{\mu \nu }^{\rho }\equiv 0$ (for generic values
of $M_{i}$ and $M_{ij}$) and return the Hilbert action as before.

From our viewpoint, $\Omega _{\mu \nu }^{\rho }$ is just a particular
higher-spin tensor multiplet, $S_{M}(g,\Omega )$ is its most general mass
term, while $S_{\mathrm{P}}(g,\Omega )-S_{\mathrm{H}}(g)$ is a particular
mass term. It is natural to view (\ref{sp}) as the low-energy limit of a
theory where $\Omega _{\mu \nu }^{\rho }$ is dynamical. The Palatini version
of quantum gravity with fakeons is the appropriate ultraviolet completion of
(\ref{sp}), or (\ref{pala}). The masses of the Palatini multiplet do not
need to be equal, or close to, the Planck mass $M_{\mathrm{Pl}}^{2}$,
because the actions (\ref{sp}) and (\ref{spp}) do not provide a
normalization for $\Omega _{\mu \nu }^{\rho }$.

Let us switch to quantum gravity with fakeons \cite{LWgrav}, which is built
starting from the action%
\begin{equation}
S_{\mathrm{QG}}(g)=-\frac{M_{\mathrm{Pl}}^{2}}{16\pi }\int \mathrm{d}^{4}x%
\sqrt{-g}\left( R+\frac{1}{2m_{\chi }^{2}}C_{\mu \nu \rho \sigma }C^{\mu \nu
\rho \sigma }-\frac{R^{2}}{6m_{\phi }^{2}}\right) ,  \label{SQG}
\end{equation}%
where $C_{\phantom{\mu}\nu \rho \sigma }^{\mu }$ is the Weyl tensor. The
cosmological term is omitted for simplicity. It is understood that the
connection $\Gamma _{\mu \nu }^{\rho }$ in (\ref{SQG}) is the Levi-Civita
one. The signs of the coefficients of $C^{2}$ and $R^{2}$ are due to the
no-tachyon condition.

Analyzing the poles of the free propagator around flat space, it is possible
to show that the theory propagates a triplet made of the graviton (that is
to say the fluctuation of the metric tensor), a scalar field $\phi $ of mass 
$m_{\phi }$ and a spin-2 fakeon $\chi _{\mu \nu }$ of mass $m_{\chi }$. Once
the cosmological term is reinstated, the theory is renormalizable. Because
the fakeon quantization prescription is adopted for $\chi _{\mu \nu }$, the
theory is also unitary \cite{LWgrav}. The scalar $\phi $ can be quantized
either as a physical particle or a fakeon.

The Palatini version of quantum gravity with fakeons is obtained by turning $%
\Gamma _{\mu \nu }^{\rho }$ into an independent field, adding all the terms
that are power-counting renormalizable, imposing that no tachyons are
present in the expansion around flat space, and finally choosing the
appropriate quantization prescriptions to have unitarity. We want to show
that theory we obtain is equivalent to (\ref{SQG}) coupled to the
(propagating) Palatini massive multiplet $\Omega _{\mu \nu }^{\rho }$.

First, we need to replace the curvature tensors $R_{\mu \nu \rho \sigma
}(g,\Gamma (g))$ of (\ref{SQG}) with $R_{\mu \nu \rho \sigma }(g,\Gamma )$,
which has reduced symmetry properties. We also need to include all the
inequivalent contractions. The list of possibilities is quite long, but some
shortcuts allow us to get to the final answer quite rapidly. The trick is to
use the variables $g_{\mu \nu }$ and $\Omega _{\mu \nu }^{\rho }$, which are
easier to handle. A further simplification suggested by (\ref{spp}) is to
assume invariance under the $\mathbb{Z}_{2}$ transformation $\Omega _{\mu
\nu }^{\rho }\rightarrow -\Omega _{\mu \nu }^{\rho }$ (we can always add the
other terms later). We can also assume parity invariance, which spares us
from using the Levi-Civita symbol $\varepsilon ^{\mu \nu \rho \sigma }$
(again, we can add the remaining terms later). Finally, we keep working in
the metric formalism, but the analysis is basically identical in the tetrad
formalism. In particular, the tetrad formalism is necessary when we want to
couple the theory to the standard model.

At this point, we have to equip (\ref{spp}) with all the two-derivative
kinetic terms we can build for $\Omega _{\mu \nu }^{\rho }$ and include the
nonminimal couplings and the $\Omega $ self interactions. We obtain 
\begin{equation*}
S_{\mathrm{QG}}^{\mathrm{P}}(g,\Omega )=S_{\mathrm{QG}}(g)+S_{\mathrm{kin}%
}(g,\Omega )+S_{M}(g,\Omega )+S_{4}(g,\Omega ),
\end{equation*}%
where%
\begin{eqnarray*}
S_{\mathrm{kin}}(g,\Omega ) &\equiv &-\frac{M_{\mathrm{Pl}}^{2}}{16\pi }\int 
\mathrm{d}^{4}x\sqrt{-g}\left[ D_{\sigma }\Omega _{\mu \nu \rho
}\sum_{i}N_{i}D^{\{\sigma }\Omega ^{\mu \nu \rho \}_{i}}+D_{\mu }D_{\nu
}\Omega _{\rho }^{i}\sum_{i}N_{i}^{\prime }\Omega ^{\{\mu \nu \rho
\}_{i}}\right. \\
&&\left. \qquad \qquad +D_{\mu }\Omega _{\nu }^{i}\sum_{i,j}\left(
N_{ij}D^{\mu }\Omega ^{j\nu }+N_{ij}^{\prime }D^{\nu }\Omega ^{j\mu }\right)
+\mathrm{nonminimal\ terms\ }\right] ,
\end{eqnarray*}%
up to total derivatives, and $S_{4}$ collects the most general quartic terms:%
\begin{equation*}
S_{V}(g,\Omega )=\int \mathrm{d}^{4}x\sqrt{-g}V^{\mu _{1}\nu _{1}\rho
_{1}\mu _{2}\nu _{2}\rho _{2}\mu _{3}\nu _{3}\rho _{3}\mu _{4}\nu _{4}\rho
_{4}}\Omega _{\mu _{1}\nu _{1}\rho _{1}}\Omega _{\mu _{2}\nu _{2}\rho
_{2}}\Omega _{\mu _{3}\nu _{3}\rho _{3}}\Omega _{\mu _{4}\nu _{4}\rho _{4}},
\end{equation*}%
where $V^{\cdots }$ are constants. Renormalizability is evident, by power
counting. On the other hand, it is nontrivial to work out the no-tachyon
conditions and the prescriptions for unitarity. For this reason, we split
the problem into simpler problems by decomposing the Palatini multiplet into
irreducible submultiplets.

\section{The Palatini multiplet}

\label{palamul}\setcounter{equation}{0}

In this section we discuss the Palatini multiplet and its main
submultiplets: the metric incompatibility submultiplet and the torsion
submultiplet.

A tensor $\Omega _{\mu \nu \rho }$ of order 3 has 64 independent components.
We are going to show that it contains one spin 3 particle, 5 spin 2
particles, 9 spin 1 particles and 5 spin 0 particles:%
\begin{equation}
\mathbf{64}=\mathbf{3}\oplus \mathbf{2}^{5}\oplus \mathbf{1}^{9}\oplus 
\mathbf{0}^{5}.  \label{64}
\end{equation}%
We start from the usual decomposition by means of Young tableaux,%
\begin{equation}
\mathbf{64}=\scalebox{0.9}[0.6]{\begin{tabular}{|l|l|l|} \hline & & \\
\hline \end{tabular}}\oplus \scalebox{0.9}[0.6]{\begin{tabular}{|l|l} \hline
& \multicolumn{1}{|l|}{} \\ \hline & \\ \cline{1-1} \end{tabular}}\oplus %
\scalebox{0.9}[0.6]{\begin{tabular}{|l|l} \hline & \multicolumn{1}{|l|}{} \\
\hline & \\ \cline{1-1} \end{tabular}}\oplus \scalebox{0.9}[0.6]{%
\begin{tabular}{|l|} \hline \\ \hline \\ \hline \\ \hline \end{tabular}}.
\label{venti}
\end{equation}

The completely symmetric part belongs to the metric incompatibility
submultiplet (see below). It has dimension 20 and can be further decomposed
into the traceless part, which is the dimension 16 multiplet $\chi _{\mu \nu
\rho }$ with free Lagrangian (\ref{l2}), and the trace part, which is a
vector multiplet with free Lagrangian (\ref{l1}): 
\begin{equation}
\scalebox{0.9}[0.6]{\begin{tabular}{|l|l|l|} \hline & & \\ \hline
\end{tabular}}\equiv \mathbf{20}_{a}=[\mathbf{3}\oplus \mathbf{2}\oplus 
\mathbf{1}\oplus \mathbf{0}]_{a}\oplus \lbrack \mathbf{1}\oplus \mathbf{0]}%
_{a}.  \label{20a}
\end{equation}%
The irreducible multiplets are distinguished by means of square brackets.

The completely antisymmetric part of $\Omega _{\mu \nu \rho }$ belongs to
the torsion submultiplet. It is a vector multiplet, by Hodge dualization,%
\begin{equation}
\scalebox{0.9}[0.6]{\begin{tabular}{|l|} \hline \\ \hline \\ \hline \\
\hline \end{tabular}}\equiv \mathbf{4}_{d}=[\mathbf{1}\oplus \mathbf{0]}_{d}%
\mathbf{,}  \label{4}
\end{equation}%
and its free Lagrangian is also (\ref{l1}).

The rest of (\ref{venti}), 
\begin{equation}
\scalebox{0.9}[0.6]{\begin{tabular}{|l|l} \hline & \multicolumn{1}{|l|}{} \\
\hline & \\ \cline{1-1} \end{tabular}}\oplus \scalebox{0.9}[0.6]{%
\begin{tabular}{|l|l} \hline & \multicolumn{1}{|l|}{} \\ \hline & \\
\cline{1-1} \end{tabular}}=\mathbf{20}_{b}\oplus \mathbf{20}_{c},
\label{2020}
\end{equation}%
contains 40 degrees of freedom, which can be split into two tensors of
dimension 20. Each of them can be further split into a traceless multiplet
of dimension 16 and a further vector multiplet: 
\begin{equation}
\mathbf{20}_{b,c}\mathbf{=}[\mathbf{2}\oplus \mathbf{2}\oplus \mathbf{1}%
\oplus \mathbf{1}]_{b,c}\oplus \lbrack \mathbf{1}\oplus \mathbf{0]}_{b,c}.
\label{20}
\end{equation}

The traceless multiplets, which decompose as $\mathbf{2}\oplus \mathbf{2}%
\oplus \mathbf{1}\oplus \mathbf{1}$ (the proof being given below), are the
only ones we have not met so far. They can be organized in different ways.
For example, we can choose $\mathbf{20}_{b}$ and $\mathbf{20}_{c}$ to
correspond to a Young tableaux of (\ref{20}) each. Or, we can choose $%
\mathbf{20}_{b}$ to belong to the metric incompatibility submultiplet. Or,
we can have $\mathbf{20}_{c}$ belong to the torsion submultiplet. Note that
we cannot satisfy the last two conditions at the same time, since the two
submultiplets mix in a nontrivial way.

To make these issues clear, we organize the right-hand side of (\ref{2020})
in two ways, which we denote by $\mathbf{20}_{b}\oplus \mathbf{20}_{c}$ and $%
\mathbf{20}_{b}^{\ast }\oplus \mathbf{20}_{c}^{\ast }$, respectively.

\subsection{Torsion decomposition}

We choose $\mathbf{20}_{b}$ to be symmetric under the exchange of $\mu $ and 
$\nu $ and $\mathbf{20}_{c}$ to be antisymmetric under the exchange of the
same indices. Then, the torsion submultiplet of $\Omega _{\mu \nu }^{\rho }$
is $\mathbf{20}_{c}\oplus \mathbf{4}_{d}$. Indeed, when $\mathbf{20}_{c}$
and $\mathbf{4}_{d}$ vanish, $\Omega _{\mu \nu \rho }$ is symmetric in the
exchange of $\mu $ and $\nu $, so $\Gamma _{\mu \nu }^{\rho }$ is
torsionless (but not necessarily metric compatible).

We denote the traceless parts of $\mathbf{20}_{b,c}$ by $\omega _{\mu \nu
\rho }^{+}$ and $\omega _{\mu \nu \rho }^{-}$, respectively. Their
projectors are%
\begin{equation*}
P_{\mu \nu \rho ,\alpha \beta \gamma }^{\pm }=\frac{1}{6}\left[ 2\eta _{\mu
\alpha }\eta _{\nu \beta }\eta _{\rho \gamma }-\eta _{\mu \gamma }\eta _{\nu
\alpha }\eta _{\rho \beta }-\eta _{\mu \beta }\eta _{\nu \gamma }\eta _{\rho
\alpha }\pm (\mu \leftrightarrow \nu )-\mathrm{traces}\right] ,
\end{equation*}%
so the desired multiplets are $\omega _{\mu \nu \rho }^{\pm }=P_{\mu \nu
\rho ,\alpha \beta \gamma }^{\pm }\Omega ^{\alpha \beta \gamma }$. A useful
property is that $P_{\mu \nu \rho ,\alpha \beta \gamma }^{\pm }$ are
invariant under cyclic permutations of $\mu \nu \rho $ and $\alpha \beta
\gamma $, so $\omega _{\mu \nu \rho }^{\pm }$ satisfy the \textquotedblleft
Bianchi identity\textquotedblright 
\begin{equation}
\omega _{\mu \nu \rho }^{\pm }+\omega _{\rho \mu \nu }^{\pm }+\omega _{\nu
\rho \mu }^{\pm }=0.  \label{bianchi}
\end{equation}%
Both multiplets $\omega _{\mu \nu \rho }^{\pm }$ contain two spin-2
particles (one physical, one fake) and two spin-1 particles (one physical,
one fake).

We focus on $\omega _{\mu \nu \rho }^{-}$, since the case of $\omega _{\mu
\nu \rho }^{+}$ is similar. Trying out all contractions and using the
Bianchi identity (\ref{bianchi}), it can be shown that the most general
quadratic, two-derivative Lagrangian for $\omega _{\mu \nu \rho }^{-}$ is%
\begin{equation}
\mathcal{L}^{-}=-\frac{1}{2}(\partial _{\sigma }\omega _{\mu \nu \rho
}^{-})(\partial ^{\sigma }\omega ^{-\mu \nu \rho })-\frac{a}{2}(\partial
^{\mu }\omega _{\mu \nu \rho }^{-})(\partial _{\sigma }\omega ^{-\sigma \nu
\rho })-\frac{a^{\prime }}{2}(\partial ^{\mu }\omega _{\mu \nu \rho
}^{-})(\partial _{\sigma }\omega ^{-\sigma \rho \nu })+\frac{m^{2}}{2}\omega
_{\mu \nu \rho }^{-}\omega ^{-\mu \nu \rho }.  \label{lmeno}
\end{equation}%
Once we work out the propagator (which we do not write here due to its
involved structure), we find the no-tachyon conditions%
\begin{equation}
a>-3,\qquad -2-a<a^{\prime }<6+a.  \label{assi}
\end{equation}%
The mass spectrum is%
\begin{eqnarray*}
\mathrm{physical\ spin\ 2\ } &:&\ m^{2}\mathrm{,}\qquad \qquad \qquad 
\mathrm{fake\ spin\ 2\ }:\ \frac{2m^{2}}{2+a+a^{\prime }}, \\
\mathrm{physical\ spin\ 1\ } &:&\ \frac{3m^{2}}{3+a}\mathrm{,\qquad \qquad
fake\ spin\ 1\ }:\ \frac{6m^{2}}{6+a-a^{\prime }}.
\end{eqnarray*}%
The nature and spin of each particle are established by investigating the
residues at the poles in the rest frame. The spectrum of $\omega _{\mu \nu
\rho }^{+}$ has an analogous structure.

\subsection{Metric incompatibility decomposition}

The second decomposition of (\ref{2020}) is $\mathbf{20}_{b}^{\ast }\oplus 
\mathbf{20}_{c}^{\ast }$, where $\mathbf{20}_{b}^{\ast }$ and $\mathbf{20}%
_{c}^{\ast }$ are symmetric and antisymmetric under the exchange of $\nu $
and $\rho $, respectively. The metric incompatibility submultiplet of $%
\Omega _{\mu \nu\rho }$ is $\mathbf{20}_{a}\oplus \mathbf{20}_{b}^{\ast } $.
Indeed, when $\mathbf{20}_{a}$ and $\mathbf{20}_{b}^{\ast }$ vanish, $\Omega
_{\mu \nu \rho }$ is antisymmetric in the exchange of $\nu $ and $\rho $, so
the connection $\Gamma _{\mu \nu }^{\rho }$ of (\ref{diffa}) is metric
compatible (but not necessarily torsionless). The Lagrangians, spectra and
no-tachyon conditions for the traceless parts $\omega _{\mu \nu \rho }^{\ast
\pm }$ of $\mathbf{20}_{b,c}^{\ast }$ are similar to the ones of $\omega
_{\mu \nu \rho }^{\pm }$. Note that when $\Omega _{\mu \nu \rho }$ is both
symmetric in $\mu \nu $ and antisymmetric in $\nu \rho $, it vanishes, so $%
\Gamma _{\mu \nu }^{\rho }$ coincides with the Levi-Civita connection.

\bigskip

The results we have found allow us to prove that the space of parameters
contains a region where all the no-tachyon conditions are fulfilled. In the
diagonal case, where the irreducible submultiplets are independent from one
another, it is sufficient to satisfy the no-tachyon conditions (\ref{as})
and (\ref{assi}) for each submultiplet separately. When the off-diagonal
mixing terms are turned on, we have situations similar to those described in
section \ref{examples}. The no-tachyon conditions continue to hold, if the
coefficients of the mixing terms are sufficiently small. As far as the
quantization prescriptions are concerned, the ``minimal'' unitary scenario
is obtained by quantizing the whole Palatini multiplet as a multiplet of
fakeons. More generally, the parameter space is divided into regions with
different contents of physical and fake particles. The spin-3 particle of (%
\ref{64}) cannot mix with other particles, so it can be quantized as a
physical particle if the right sign is chosen in front of its kinetic term.
The particles that have the same spin belonging to the list (\ref{64}) mix
with one another and, depending on the features of the mixing, they may be
quantized as physical particles or not. What is important is that we have
proved that there are quantization prescriptions and a region in parameter
space that make the theory perturbatively unitary.

In conclusion, the Palatini version of quantum gravity with fakeons is just
the non-Palatini one coupled to the Palatini multiplet. Consequently, it is
equally renormalizable and unitary, once the no tachyon conditions are
fulfilled and the right prescriptions are chosen for the various poles of
the free propagators in the expansion around flat space.

\section{Conclusions}

\label{conclusions}\setcounter{equation}{0}

Quantum field theories of massive particles of arbitrary spins can be
formulated in a local, renormalizable and unitary way by embedding the
particles into larger multiplets, containing fakeons and possibly other
physical particles.

The models built along this guideline uncover a hidden subsector of quantum
field theory. The interactions are similar to the ones of ordinary theories
of particles of lower spins, and so are the couplings to gravity and gauge
fields. A notable example is the Palatini version of quantum gravity, which
is just the non-Palatini one coupled to a peculiar multiplet of order 3.
More generally, the models provide an arena for investigating new types of
RG flows, fixed points and asymptotically safe theories. An unexpected
feature of the RG fixed points is that they are in general only scale
invariant, but not conformal invariant. Exceptions exist, if we restrict to
fermionic multiplets, make use of large $N$ expansions, concentrate on the
Euclidean versions of the theories.

Arbitrary-spin massive multiplets may provide candidates for physics beyond
the standard model, the matter sector of quantum gravity and possibly dark
matter. To be good dark-matter candidates, maybe WIMPs, they must be
electrically neutral, neutral with respect to $SU(3)$ and couple to the $Z$
boson very weakly.

Generically, we can expect the multiplets to be superheavy (that is to say,
have masses larger than 10$^{12\text{-}14}$GeV), for the following reasons.
We know that fakeons must be massive to restrict the violation of causality
to small distances. We have shown that RG invariant identities, such as (\ref%
{RGmas}) and (\ref{RGmasf}), relate the masses of the particles within an
irreducible multiplet. This property suggests that the masses of the
particles belonging to a multiplet might be comparable to one another. We
also know that the only fakeon that must necessarily exist for the
consistency of quantum gravity is the spin 2 fakeon $\chi _{\mu \nu }$ of
mass $m_{\chi }$ predicted by the theory (\ref{SQG}). As shown by Bianchi,
Piva and the current author in \cite{ABP}, its mass should satisfy the bound 
$m_{\chi }>m_{\phi }/4$ on cosmological grounds. Current data on the power
spectrum and tilt of the primordial scalar fluctuations of the CMB radiation 
\cite{Planck18} imply that the scalar field $\phi $ is superheavy. Then the
bound $m_{\chi }>m_{\phi }/4$ implies that $\chi _{\mu \nu }$ is also
superheavy. If other fakeons exist, as in the multiplets we have considered
here, maybe all of them have the same origin (not known today) and
comparable masses. If that is the case, the whole multiplets might be
superheavy. As discussed by Chung, Crotty, Kolb and Riotto in refs. \cite%
{riotto} superheavy dark matter is a viable option.

On the other hand, at present we cannot completely exclude the relevance of
light multiplets for cosmology and astroparticle physics. Light spin-2
(Pauli-Fierz) dark matter candidates have been considered in ref. \cite%
{raidal} by Marzola, Raidal and Urban in the context of bimetric gravity 
\cite{bigravity}. The multiplets studied here might be worth of attention in
these kinds of searches.

\vskip12truept \noindent {\large \textbf{Acknowledgments}}

\vskip2truept

I am grateful to D. Comelli, C. Iazeolla and A. Strumia for useful
correspondence.

\end{document}